OPTICAL MANAGEMENT TECHNIQUES FOR ORGANIC SOLAR CELLS

By

ADHARSH RAJAGOPAL

A THESIS PRESENTED TO THE GRADUATE SCHOOL
OF THE UNIVERSITY OF FLORIDA IN PARTIAL FULFILLMENT
OF THE REQUIREMENTS FOR THE DEGREE OF
MASTER OF SCIENCE

UNIVERSITY OF FLORIDA

2014

To my family

ACKNOWLEDGMENTS

This work was made possible with the kind support and help of many individuals. First, I extend gratitude to my graduate advisor, Dr. Jiangeng Xue for providing me the opportunity to work in his research group. It has been a great learning experience and I enjoyed working in his group. His valuable guidance and high expectations has facilitated the development of my research skills and interests. I also acknowledge my other committee members Dr. Franky So and Dr. Jennifer Andrew for their time and interest.

My labmates have played a key role in successful completion of my research work. I would like to thank Dr. Weiran Cao for continuously monitoring the progress of my work. His timely suggestions and ideas guided me to progress faster. I am also thankful to John Mudrick and Nathan Shewmon for their support and help right from the initial lab training till the end of my work. I also thank Chenchen Yang for being a good companion in the lab.

My research has been made possible by the financial support from National Science Foundation, Major Research Instrumentation program, Research Corporation for Science Advancement Scialog program, Department of Energy Solar Energy Technologies Program and University of Florida Office of Science

Finally, yet importantly I would like to express gratitude to my family and friends. My parents and sister provided me consistent support with their words of encouragement. Their belief about my abilities helped me to overcome the hardships that I faced. This thesis is dedicated to them.

# TABLE OF CONTENTS

LIST OF TABLES

# LIST OF FIGURES

LIST OF ABBREVIATIONS

| | |
|---|---|
| AFM | Atomic Force Microscope |
| Ag | Silver |
| Al | Aluminum |
| AM 1.5G | Air Mass 1.5 Global |
| AR | Anti-reflective |
| ASTM | American Society for Testing and Materials |
| BHJ | Bulk Heterojunction |
| c | Speed of light |
| DI | Deionized |
| EQE | External Quantum Efficiency |
| f | Enhancement factor |
| FDTD | Finite difference time – domain |
| FF | Fill Factor |
| h | Planck's constant |
| HJ | Heterojunction |
| HOMO | Highest Occupied Molecular Orbital |
| $I_D$ | Incident current measured by silicon photo detector |
| IPA | Isopropyl alcohol |
| IQE | Internal Quantum Efficiency |
| $I_{SC}$ | Short-circuit current |
| $I_T$ | Photocurrent of test cell |
| ITO | Indium Tin Oxide |
| I-V | Current-Voltage |
| $J_{SC}$ | Short-circuit current density |
| LB | Langmuir-Blodgett |
| LSCs | Luminescent solar concentrators |
| LUMO | Lowest Unoccupied Molecular Orbital |
| M | Spectral mismatch factor |
| MLA | Microlens Array |
| $MoO_x$ | Molybdenum oxide |
| MOs | Molecular Orbitals |
| n | Refractive index |
| NOA | Norland Optical Adhesive |
| OLEDs | Organic Light Emitting Diodes |
| OMO | Oxide/Metal/Oxide |
| OPVs | Organic Photovoltaics |
| OSC | Organic Solar Cell |
| P3HT | Poly(3-hexylthiophene-2,5-diyl) |
| PCBM | Phenyl-C61-butyric acid methyl ester |
| eV | Electron-volt |
| PCE (or) $\eta_P$ | Power Conversion Efficiency |
| PDMS | Poly(dimethylsiloxane) |
| PEDOT:PSS | Poly(3,4-ethylenedioxythiophene):Polystyrene sulfonate |
| PHJ | Planar Heterojunction |
| $P_{in}$ | Incident power |
| $P_{max}$ (or) $P_{m,out}$ | Maximum power output |
| PM-HJ | Planar Mixed Heterojunction |
| $P_o$ | Incident power |
| PR | Planar Reflector |
| PV | Photovoltaic |
| q | Charge of electron |
| QTH | Quartz Tungsten Halogen |
| R | Reflector |
| $R_D$ | Responsivity of detector |
| rpm | Rotations per minute |
| $S(\lambda)$ | Reference spectrum |
| SEM | Scanning Electron Microscope |
| $SiO_2$ | Silica |
| ST | Semi-transparent |
| SubPc | Subphthalocyanine chloride |
| t | Thickness of active layer |
| TR | Textured Reflector |
| UV | Ultraviolet |
| $V_{OC}$ | Open-circuit voltage |
| VTE | Vacuum Thermal Evaporation |
| Xe | Xenon |
| ZnO | Zinc Oxide |
| $\eta_A$ | Absorption efficiency |
| $\eta_{CC}$ | Charge collection efficiency |
| $\eta_{CT}$ | Charge transfer efficiency |
| $\eta_{ED}$ | Exciton diffusion efficiency |
| $\Theta_C$ | Critical angle |
| $\lambda$ | Wavelength |

Abstract of Thesis Presented to the Graduate School
of the University of Florida in Partial Fulfillment of the
Requirements for the Degree of Master of Science

OPTICAL MANAGEMENT TECHNIQUES FOR ORGANIC SOLAR CELLS

By

Adharsh Rajagopal

August 2014

Chair: Jiangeng Xue
Major: Materials Science and Engineering

Over the last two decades, the field of organic photovoltaics (OPVs) has started gaining increased attention among researchers in academic and industrial communities. Consistent efforts are being taken to increase the maximum attainable efficiency for OPV technology through development of new materials and device structures. However there have been very few studies related to optical management, which is a promising approach to improve device performance in OPVs. It involves manipulating the light path in order to achieve an increased photon flux within the active layer. This approach improves the optical absorption of the active layer leading to an increase in short-circuit current and power conversion efficiency for OPVs. In this thesis, two different optical management techniques for organics based solar cells are explored.

The first part is focused on the development of a textured rear reflector for OPVs. The use of textured reflector (TR) facilitates an increase in the optical path length along with light trapping within the active layer. TR was fabricated through a relatively simpler technique by depositing metal films over a microlens array (MLA). Zinc oxide nanoparticles were used to minimize the shadowing effect. Using TR, enhancements in short-circuit current density and power conversion efficiencies up to 10-25% were demonstrated for a polymer based organic solar cell.

The second part is focused on improving the effectiveness of MLA incorporation in OPVs. The increase in path length achieved using MLA can be improved by increasing the refractive index of MLA and incorporating MLA directly on the transparent electrode instead of glass substrate. This approach could avoid the optical losses occurring at the interface between MLA and glass substrate and increase the net path length of light within the active layer. Initial experimental results are discussed to understand the potential of this approach. Approximately, ~15% enhancement in short-circuit current density was achieved by using MLA directly on the transparent electrode in ITO-free devices.

CHAPTER 1

FUNDAMENTALS OF ORGANIC PHOTOVOLTAIC DEVICES

**1.1 Introduction**

Photovoltaic (PV) technology involves conversion of sunlight into electricity and is a promising source for clean and sustainable form of energy. It is expected to play a major role in meeting energy needs of the future[1,2]. Currently, the best performing solar cells in the market have active areas composed of inorganic semiconductors such as silicon and gallium arsenide. The production of inorganic solar cells has been fine-tuned over several decades and a record efficiency of 44.7% has been achieved[3]. However many fabrication steps still requires high temperature processing with modest throughput rates and involves expensive raw materials[4]. The cost per watt of electricity generated is very high and thus impeding the wide-spread commercialization of solar cells as a main form of energy[5].

Over the last two decades, the field of organic photovoltaics (OPV) has started gaining increased attention among researchers in academic and industrial communities. Since the first demonstration of 1% efficient organic solar cells by C.W. Tang in 1986[6], an exponential increase in research is observed for PV technology based on organic materials[7]. The gaining popularity and commercialization of organic light emitting diodes (OLEDs) has started boosting the growth of OPVs. Owing to their organic active layer, OPV devices offer the promise of mechanical flexibility for roll-to-roll printing, low temperature fabrication methods and lightweight devices[8]. Further, based on organic chemistry the properties of materials composing the active layers can be fine-tuned by a vast array of synthesis methods[9]. The current state of the art OPV device efficiencies[10] have reached ~11%. The reduced manufacturing and materials costs for OPVs suggest that, electricity from organic solar panels with efficiencies comparable to that of inorganic modules can be produced at a relatively lower cost. Consistent research efforts are required to improve the device performance and stability of OPVs for facilitating the commercialization of this technology[11].

In this chapter, the key concepts related to organic solar cells are reviewed in order to provide the reader a fundamental understanding of OPV technology. This is very essential to comprehend the work presented in the subsequent chapters of this thesis. Section 1.2 provides a brief overview of organics semiconductors. The mechanism for photoelectric conversion in OPVs is discussed Section 1.3. The important parameters in organic solar cells and the approaches to achieve higher efficiencies in OPVs are discussed in Section 1.4 and 1.5 respectively. The importance of transparent electrodes in OPV devices is explained in Section 1.6 (because of its relevancy to the work presented in Chapter 3 and 4). The details about device fabrication and characterization are provided in Section 1.7 and 1.8 respectively. Finally an outline of the subsequent chapters in this thesis is given in Section 1.9.

## 1.2 Overview of Organic Semiconductors

### 1.2.1 Introduction to Organic Solids

Small Molecules and Polymers are the two most important classes of organic materials with semi-conducting properties. Small molecules are the simplest form of organic materials, with masses of several hundreds of atomic mass unit. Whereas, polymers are more complex with long chain of repeating units and masses ranging up to millions of atomic mass unit. Molecular structures of organic molecules used in this work (P3HT and PCBM) are given in Figure 1-1. The interactions of weaker parallel $p_z$ orbitals gives rise to weaker bonding (π) and anti-bonding (π*) molecular orbitals (MOs). Since π- π* transition is the lowest energy option for electronic excitation in organic molecules, π-bonding MO are referred to as "highest occupied molecular orbital (HOMO)" and π*-antibonding MO are referred to as "lowest unoccupied molecular orbital (LUMO). The HOMO and LUMO are corresponding to the valence band and conduction band in inorganic solids.

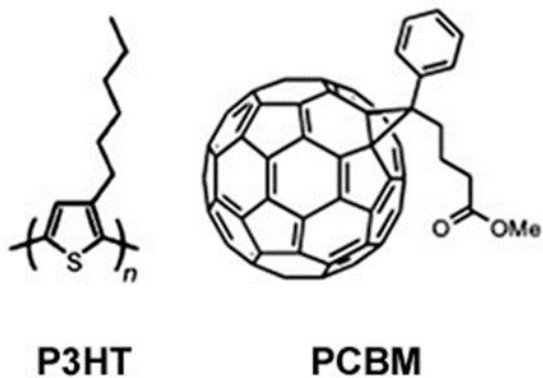

Figure 1-1. Molecular structures of P3HT (Poly (3-hexylthiophene-2, 5-diyl)) and PCBM (Phenyl-C61-butyric acid methyl ester).

**1.2.2 Charge Transport in Organic Semiconductors**

Organic semiconductors are π-conjugated molecules which contains alternating carbon-containing single and double bonds in their molecular structure. The electronic properties are determined by the degree of π-conjugation. Greater conjugation length leads to greater degree of delocalization of electrons and in turn high mobility of charge carriers. However, the mobility in organic molecules are greatly limited because of the weaker intermolecular van der walls forces[9]. This leads to high degree of charge localization within a molecule and prevents continuous band transport as in inorganic molecules[12]. The intermolecular transport in organic molecules generally occurs through a hoping process and has an energy barrier associated with it.

**1.2.3 Excitons**

In optoelectronic devices based on organic molecules, the charge localization leads to a bound electron-hole pair known as exciton. Exciton are chargeless and can diffuse through band transport or hopping processes. The weaker interactions limits the exciton mobility and diffusion lengths in organic solids[12]. Strong excitonic bonding energy (~1-2 eV) bounds the charge carriers (electrons and holes) and prevent the free movement of charge carriers. This creates a fundamental difference in the operation of organic and inorganic semiconductors. The dissociation of excitons in organic semiconductors are generally done by introduction of a heterojunction with the HOMO and LUMO level offsets greater than the excitonic binding energy[6]. On reaching such an interface, the exciton dissociation into free charge carriers becomes energetically favorable.

**1.3 Mechanism for Conversion of Light into Electricity in OPVs**

The operation of organic photovoltaics are based on donor-acceptor heterojunction with suitable energy level offsets. The process of conversion of photon into electron-hole pairs to generate electricity by OPVs can be divided into four stages[7]. The process starts with the absorption of photon within organic molecules. This leads to generation of excitons in donor and acceptor molecules. Then, the excitons diffuses to reach the heterojunction. In organic molecules, the typical exciton diffusion length is ~10 nm, beyond which annihilation of bound electron-hole pair occurs. Once the excitons reach the heterojunction, dissociation of exciton occurs through a rapid charge transfer process leading to the transfer of holes to the HOMO of donor material and electrons to the LUMO of acceptor material. After this, due to the built-in electric field and the concentration gradient of the carriers, the electrons and holes are transported through the acceptor and donor molecules to cathode and anode respectively. Once the carriers are collected at the respective electrodes, photocurrent is generated. The limited exciton diffusion and charge collection lengths for organics is the biggest challenge in development of organic solar cells. The entire process of photo-electric conversion is schematically described in Figure 1-2.

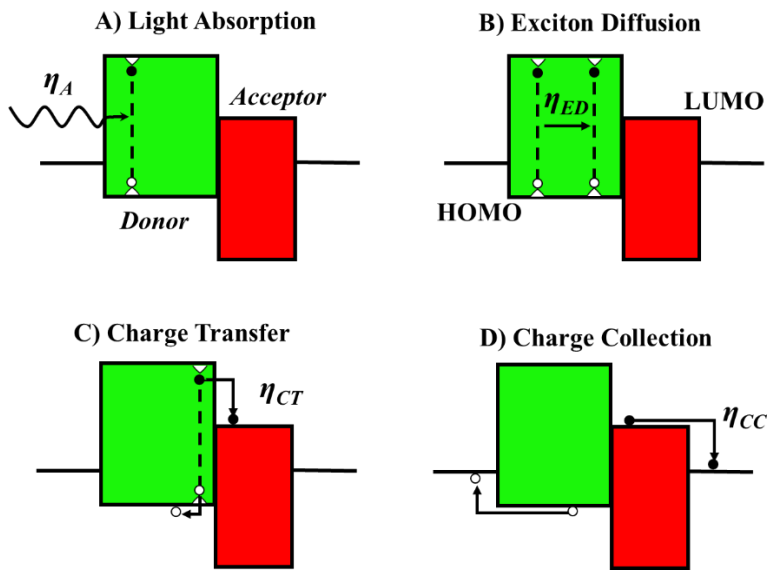

Figure 1-2. Schematic of four steps in the photovoltaic process for an organic donor-acceptor heterojunction. A) Light Absorption. B) Exciton Diffusion. C) Charge Transfer. D) Charge Collection.

### 1.4 Important Parameters of an Organic Solar Cell

The most common way to characterize a solar cell is the measurement of current-voltage characteristics. The figures-of-merit are short-circuit current ($J_{SC}$), open-circuit voltage ($V_{OC}$) and the fill factor (FF). Figure 1-3 shows the current-voltage characteristics (I-V) for an organic solar cell under dark and illumination. Important parameters are indicated in Figure 1-3. Short-circuit current and open-circuit voltage can be defined as the current at zero voltage and the voltage at zero current respectively. In order to compare the performance of devices with different areas, it is a common practice to convert current (*I*) into current density (*J*) using *J* = *I* / *A*, where *A* is the device area. Fill factor (FF) is related to the maximum power output (*P*max) and can be defined by: $FF = P_{max}/J_{SC} \cdot V_{OC}$, where *P*max is the maximum product of current and voltage across the entire J-V curve. The power conversion efficiency (*η*P) is given by the ratio of maximum power output (*P*m,out) to the incident power (*P*in). i.e, $\eta_P = P_{max}/P_{in}$.

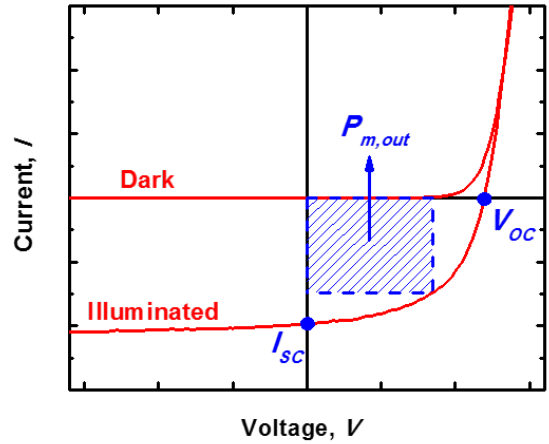

Figure 1-3. Schematic illustration of an I-V curve for organic solar cell under dark and illumination. The power conversion efficiency is calculated as the ratio of *P*m,out to the incident power. *I*SC and *V*OC are short-circuit and open-circuit voltage respectively.

Quantum efficiencies are important to understand the spectral behavior of OPVs. There are two types of quantum efficiency: the external quantum efficiency (EQE), a ratio of the number of electrons collected per number of photons incident on the device, and the internal quantum efficiency (IQE), a ratio of the number of electrons collected per absorbed photon. EQE can be calculated using the product of efficiencies of four fundamental processes in organics solar cells and is given by: $\eta_{EQE} = \eta_A \cdot \eta_{ED} \cdot \eta_{CT} \cdot \eta_{CC}$, where *η*A, *η*ED, *η*CT and *η*CC represent the efficiencies of light absorption, exciton diffusion, charge transfer and charge collection. The EQE and IQE are related as *η*EQE = *η*A × *η*IQE.

## 1.5 Approaches for Efficiency Enhancement

The device efficiency, stability and mass production are the most common issues in the growth of OPV technology[13]. A proper understanding of all the factors which affect the important parameters of an organic solar cell is very essential to improve the device efficiency and performance. In this section techniques used to improve the efficiency and stability of OPV devices are discussed. Circumventing the "exciton-diffusion bottleneck"[14] is one of the most important challenge. The most commonly used approaches are: development of new active layer materials through molecular engineering, development of new device structures interfaces and optical management.

### 1.5.1 New Materials

The optical and electronic properties of organics can be greatly tuned synthetically. Development of new materials with high absorption coefficient for active layers can be used improve the efficiency of OPVs without increasing the layer thickness. Currently most organic materials have a relatively large band gap (1.5 – 2 eV) resulting in relatively low absorption. Developing low band gap organic semiconductors[15,16] is one way to further extend the light absorption into deep infrared region. Organic materials also suffer from poor charge mobility. This can be improved by attaining materials with high purity and crystalline characteristics[17,18]. Another strategy to successfully harvest the photons is to use molecular engineering to adjust the levels of HOMO and LUMO energy bands. The HOMO level should be lower to enhance the $V_{OC}$ and LUMO energy levels should be suitable for efficient electron transfer to fullerenes (which is the most commonly used acceptor molecule in OPVs)[19]. These approaches for developing new materials can greatly improve the device performance of OPVs.

**1.5.2 Device Structure and Interfaces**

The dissociation of exciton with strong binding energy requires the use of active layer with heterojunction (HJ) for OPVs. The most commonly used architectures are planar heterojunction (PHJ), bulk heterojunction (BHJ) and planar-mixed heterojunction (PM-HJ). PHJ[6] involves sandwiching planar layers of donor and acceptor materials between two electrodes. The advantage of this is that the carrier recombination is greatly avoided after charge separation. However, the layers have to be thin enough to facilitate effective exciton diffusion. BHJ[20] involves the use of blend layer of donor and acceptor materials which forms an interpenetrating network. The increased number of interfaces greatly improve the exciton diffusion. However, the charge collection is relatively less effective than PHJ. Planar-mixed HJ[21] is a combination of both PHJ and BHJ. In this architecture, the active layer consists of a mixed layer between two planar layers. With proper optimization this structure can have the advantages of both PHJ and BHJ. Apart from active layers, developing interfacial layers[22] and favorable morphologies[23] can greatly improve the charge collection at the electrodes.

Improvement in stability of OPVs can be achieved with the development of new device configurations. The commonly used normal configuration is: Glass / Anode / Active layer / Cathode. The oxidation of low work function[19] cathode causes degradation in air. This can be circumvented with the use of inverted configuration: Glass / Cathode / / Active layer / Anode. A typical example of normal and inverted configuration for a P3HT:PCBM based BHJ is shown in Figure 1-4.

Thus, these are some ways in which device structure can be modified and used to enhance the performance of OPVs.

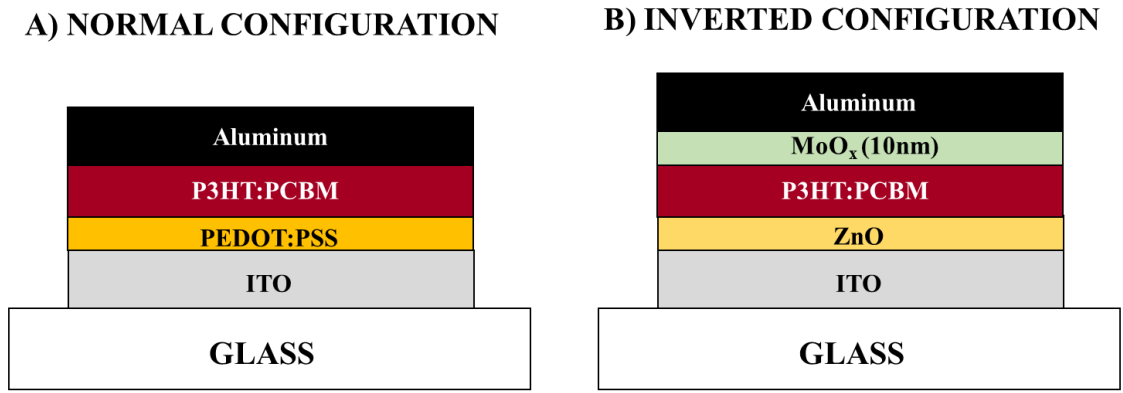

Figure 1-4. Schematic for a typical P3HT:PCBM BHJ based normal and inverted configuration. A) Normal configuration: Glass / ITO / PEDOT:PSS / P3HT:PCBM / Al. B) Inverted configuration: Glass / ITO / ZnO / P3HT:PCBM / MoO$_x$ / Al.

### 1.5.3 Optical Management

The above mentioned approaches may be time consuming and expensive for betterment of OPVs. A relatively simpler approach of manipulating the light within the device can be used to achieve enhancement in efficiencies[7]. With any existing material system and device structures, optical management techniques can be used to increase the amount of light within the active layer. This would lead to an increased light absorption and generate more charge carriers. In turn, an improved short-circuit current can be achieved. Further details about optical management are discussed in Chapter 2. The work presented in this thesis (Chapter 3, 4) are based on this approach.

### 1.6 Transparent Electrodes for OPV Devices

Development of efficient transparent conductive electrodes is essential for making progress in OPVs[24]. The transparency of electrodes determine the amount of light transmitted to the active layer. Indium Tin Oxide (ITO) is the most commonly used transparent electrode in OPVs. However due to the use of rare metal indium along expensive production process, replacements to ITO are being developed[25]. Thin metal layers[26], transparent conductive polymers[25], graphene[27], carbon nanotube[28], metal nanowires[29] are some of the alternatives which are currently being studied.

Generally, OPV device structure uses transparent electrode on one side and reflective electrode on the other side. Conversely the use of two transparent electrodes may be required at some instances. It can be very helpful in developing visually transparent OPV devices for building-integrated PV systems[30]. Most importantly, the use of two transparent electrodes is greatly useful for integration of several optical structures. Integration of pyramidal rear reflectors[31] to transparent devices is one such example. The textured rear reflector discussed in Chapter 3 also utilizes a transparent OPV device. In these cases, if the electrodes do not have sufficient transparency, they account for some of optical losses because of absorption. This reduces the amount of light available for absorption within the device and thus greatly reduces the effectiveness of the optical structures.

## 1.7 Device Fabrication Techniques

Thermal vacuum deposition and spin coating are the two approaches used for fabrication of organics solar cells in this work. Other techniques[32] like inkjet printing, spray coating and doctor blading can be used, but are not discussed here.

### 1.7.1 Thermal Vacuum Evaporation

Vacuum thermal evaporation (VTE) are generally used for depositing thin films of organic small molecules. A typical VTE system consists of source material boats, electrodes for resistive heating, quartz crystal monitor and shutter to precisely control the film thickness and a shadow mask to pattern the substrate. The source-substrate distance for the chamber used in this work is approximately 20 cm. This large distance results in less efficient use of source materials because only a small fraction of the molecular beam is incident on the substrate. But it improves the uniformity of film thicknesses across a number of substrates.

The source material is loaded in the boats and the system is evacuated to high vacuum (<3 x $10^{-6}$ torr). The boats are then resistively heated for evaporation of the source materials. During evaporation, the substrate holder is rotated (~5-6 rpm) to achieve uniform thickness. A quartz crystal microbalance is used to monitor the rate and film thickness during deposition. The films produced by VTE are of high quality. However, the divergent nature of the evaporation beam, mask thickness and separation distance of the substrate causes a shadowing effect during deposition. This results in features slightly bigger than the shadow mask pattern.

### 1.7.2 Spin Coating

Polymers cannot be deposited by VTE because their decomposition temperature is less than their evaporation temperature. Spin coating is the most common route in laboratories to produce thin films for polymer based PVs. Uniform films of defined thickness can be achieved by spin coating of a solution. Polymer molecules are dissolved in a specific solvent (depending on their solubility) and stirred overnight to achieve complete dissolution without any particulates. The spin coating process involves four stages. Initially sufficient amount of solution is dropped on the substrate held by vacuum in a spin coater. In the second stage, the substrate is accelerated to a desired spin speed. The centrifugal forces the solution towards the edge of the substrate. In the third stage, excess solution drops of the film and a uniform thickness is achieved based on the balance between the centrifugal force and the viscosity of the solution. Finally, the film is dried off. The drying process is accelerated using thermal or vacuum annealing. Despite of the inefficient material usage in spin coating, it is a good to use in laboratory for making polymer based devices.

## 1.8 Characterization of OPV Device Performance

### 1.8.1 Current-Voltage Characteristics

The current-voltage (J-V) characteristics are tested using an Agilent 4155C semiconductor parameter analyzer, which biases the test OPV and measures the output current. A schematic of the test setup is given in Figure 1-5. Devices are tested under a 100 mW/cm$^2$ (1 Sun) simulated AM 1.5 solar spectrum using an Oriel solar simulator. The solar simulator consists of a 150W Xe-arc bulb with a parabolic reflector to increase irradiance. An AM 1.5G filter is used to remove the strong UV emission of the Xe-arc bulb. Optics are used to collimate the beam such that the beam is uniform over an area of 3 cm$^2$ in the center. However, the intensity may vary towards the edges. The intensity of the simulator is measured using a calibrated single-crystalline silicon reference cell, with KG1 short pass filter in order to minimize the spectral mismatch between the P3HT:PCBM solar cells and the reference cell. The mismatch factor was calculated as 1.03 (for P3HT:PCBM BHJ) and it was taken into account to illuminate device with a true 1-sun spectrum as per ASTM E973. The intensity can be slightly adjusted by changing the lamp power (150±5 W) to account for variability in the spectral mismatch factors and age of lamp. Neutral density filter wheels are placed in between the solar simulator and the test plane to adjust the incident power from 0.1 sun to greater than 1 sun.

The devices are placed in a test pocket at the same plane as that of the silicon reference cell during calibration. The test pocket is moved such that the device is positioned in the center uniform portion of the beam from simulator. The illumination area is much greater than the device (0.04 cm$^2$) and planar / textured reflector (4 cm$^2$ base area) in order to study the complete outcome of the textured reflector (Chapter 3). Similarly, for MLA incorporation (Chapter 4) beam larger than the device area was used. For dark current measurements, a shutter is used to block the light from the simulator and the test pocket is covered with a black cloth to ensure a light-free environment. During J-V measurements, double scans are used to detect hysteresis in the device performance, if any. This can be an indicative of charge trapping or other electronic defects within the active layer. The important photovoltaic parameters (short-circuit current ($J_{SC}$), open-circuit voltage ($V_{OC}$), and the fill factor (FF)) are calculated from two individual light scans and averaged.

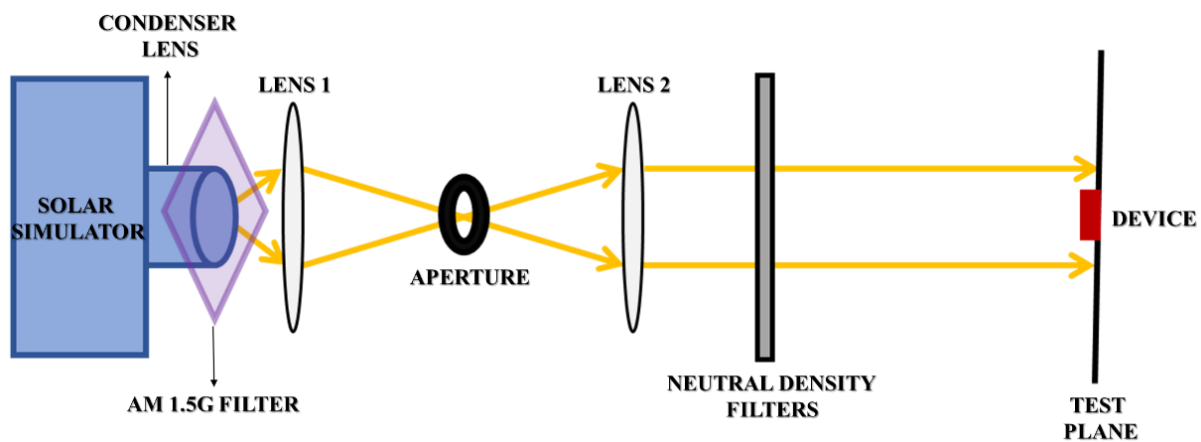

Figure 1-5. Schematic describing experimental setup for measurement of current-voltage characteristics.

### 1.8.2 External Quantum Efficiency Measurement

Quantum efficiency measurement is another vital technique to assess and compare the performance of solar cells. They are done to understand the spectral responsivity of the solar cell and calculate the integrated short-circuit current value under reference 100 mW/cm$^2$ AM 1.5G illumination. As discussed above, the EQE and IQE are related as: $\eta_{EQE} = \eta_A \times \eta_{IQE}$, where $\eta_A$ is the absorption efficiency. The IQE measurements are critical only to study the design of new material or device architectures. However for the incorporation of optical structures (Chapter 3-4), EQE is more relevant because of $\eta_A$ is affected and $\eta_{IQE}$ remains unchanged. Hence calculation of IQE is not discussed here.

The quantum efficiency measurements are made as per ASTM E1021 testing standard. A schematic for the test setup is shown in Figure 1-6. A quartz tungsten halogen (QTH) lamp feeds white light into a monochromator. Inside the monochromator, the diffraction gratings are used to split the white light into monochromatic beam of desired wavelength, which is of low intensity (1-10 µW/cm$^2$). The diverging monochromatic beam is collimated and chopped using a mechanical blade to create an alternating signal. Another set of optics is used to either collimate or focus the beam at the test plane. In case of a focused beam, the beam size is small enough to be completely placed within the device to avoid discrepancies from the device area measurement. However, collimated beam larger than the device area is used in case of measurement for devices with textured reflector or microlens array. This is essential to avoid edge effects and study the enhancement characteristics completely.

The test device is also illuminated by a non-chopped white light (0.7-1 sun intensity) in order to mimic the measurement under 1 sun using the J-V characterization system. The photocurrent is passed through a current amplifier to increase the signal/noise ratio, and then inputted into a lock-in amplifier. The lock-in amplifier uses the reference frequency from the optical chopper to isolate the current component that comes solely from the chopped monochromatic light. A calibrated Si-photo detector is placed at the test plane to measure the incident current ($I_D(\lambda)$), which is converted to incident power ($P_0(\lambda)$) by dividing with the responsivity of the detector ($R_D(\lambda)$). The EQE is calculated using $EQE(\lambda) = \frac{hc}{\lambda}\frac{I_T(\lambda)}{P_0(\lambda)}$, where h is Planck's constant, c is speed of light, $I_T(\lambda)$ is photocurrent of the solar cells measured using lock-in amplifier and $P_0(\lambda)$ is the incident power. For illumination area larger than the device, the incident power is adjusted to account only for power incident on the device area. The short-circuit current under reference 100 mW/cm² AM 1.5G illumination is calculated from the EQE data using the following integration: $J_{SC} = \int_{\lambda_1}^{\lambda_2}\frac{q\lambda}{hc}EQE(\lambda)S(\lambda)d\lambda$, where S($\lambda$) is the reference AM 1.5G spectrum.

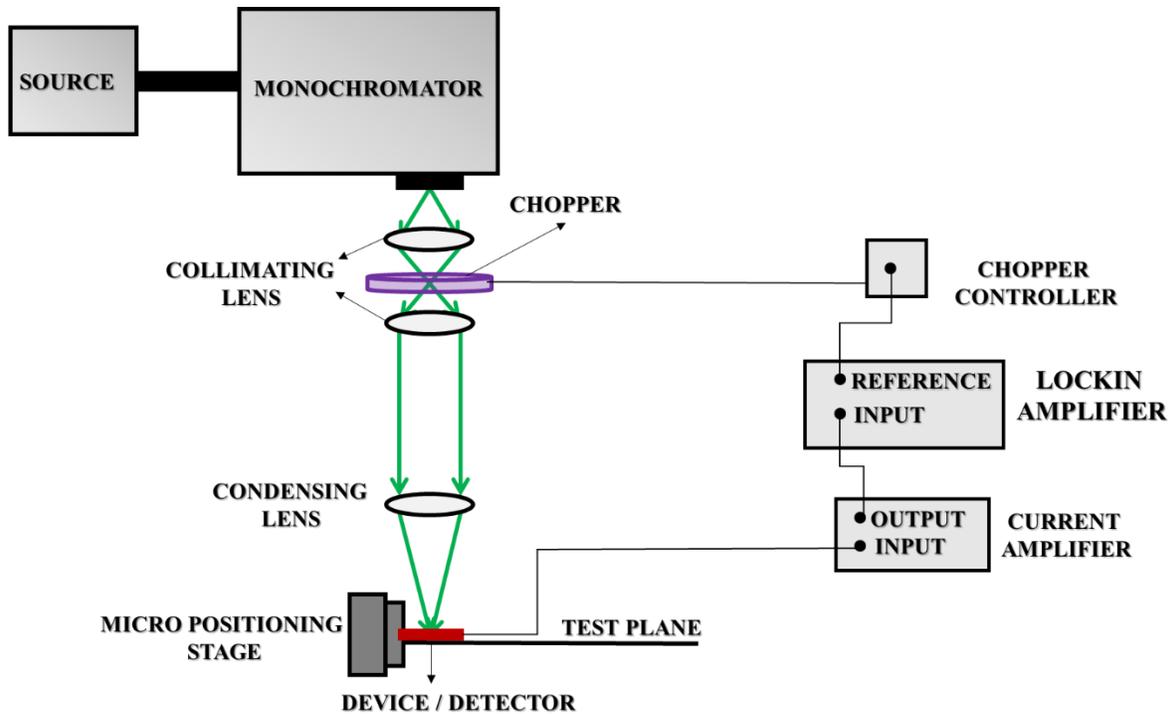

Figure 1-6. Schematic for test setup used to measure the external quantum efficiency.

### 1.9 Outline of the Thesis

The main focus of this thesis is optical management in organic solar cells. Important concepts related to organic semiconductors and organic photovoltaics along with the fabrication techniques and characterization methods for OPV are presented in Chapter 1. It provides required knowledge to the reader for following the subsequent chapters.

Chapter 2 consists a review of the reported optical management techniques for organic solar cells. It discusses about light trapping with non-planar device designs (Section 2.2), optical structures for planar devices (Section 2.3) and some other light trapping ideas (Section 2.4). A summary is provided at the end in Section 2.5. This chapter is vital to understand the context of work presented in this thesis.

Chapter 3 focusses on the development of textured reflector for organic solar cells. The design of textured reflector along with their mechanism for efficiency enhancement are discussed in Section 3.2. Experimental details about the fabrication and incorporation of textured reflector are given in Section 3.3. The results of enhancement characterization as well as the suitability of textured reflector are discussed in Section 3.4. A summary is provided at the end in Section 3.5.

Chapter 4 suggests a technique for enhancing the effectiveness of MLA incorporation in OPVs. The concept of MLA incorporation on transparent electrode is presented in Section 4.2. Experimental details are provided in Section 4.3. The results for comparison of MLA incorporation on glass and transparent electrode along with incorporation in ITO-free devices are discussed in Section 4.4. A summary is provided at the end in Section 4.5. Finally, Chapter 5 summarizes the conclusions and provides suggestions for future work.

# CHAPTER 2
# REVIEW OF OPTICAL STRUCTURES FOR ORGANIC SOLAR CELLS

## 2.1 Introduction

In organic photovoltaics (OPVs), the poor mobility of the active layer materials (ranging between ~$10^{-5}$ to 1 cm$^2$/Vs) along with the short exciton diffusion lengths (10–20 nm) necessitates the need for very thin active layer (20–250 nm) to provide effective charge collection and exciton diffusion. This in turn controls the internal quantum efficiency for OPVs. However, the thickness is too thin for optical absorption and results in significant optical losses. A substantial amount of incident light is transmitted without absorption. Based on Beer Lambert's law, the absorption efficiency can be increased exponentially by increasing the active layer thickness. This cannot be done because increased active layer thickness affect the internal quantum efficiency adversely. To achieve a tradeoff between these two factors, research is focused on the development of new materials and device architectures. Alternatively, optical management (a relatively less exploited approach) can be used to greatly improve the device performance of organic solar cells.

Optical management techniques involve the use of optical structures to achieve enhanced absorption with thin active layers. This could improve the amount of light absorbed in an organic solar cell without affecting its electrical processes. In general, organic solar cells are used in two types of configurations: substrate and superstrate configuration. A substrate configuration comprises of a transparent or reflective substrate to hold the device, with light illumination through a transparent electrode on the other side. On the other hand, a superstrate configuration involves building the device on a transparent substrate like glass such that the substrate holds the device and acts as a window for illumination. In both these configurations, the use of optical structures increases the photon flux within the active layer. The important mechanisms through which more light is absorbed within active layer are: avoiding reflection at the light incoupling surface, trapping light and increasing the mean path length of light within the active layer.

A number of reports are available in the literature for optical management in OPVs. In this chapter, some of these reported techniques are consolidated and reviewed to understand the context of the work in this thesis (Chapter 3 and 4) in a better way. Non planar device designs which are used for light trapping are discussed in Section 2.2. Some of the external optical structures incorporated with the planar devices are discussed in Section 2.3. Finally, some other light trapping ideas reported for OPVs are mentioned in Section 2.4. A summary is provided at the end in Section 2.5. Beyond the concepts discussed here, much detailed reviews of optical structures can be found elsewhere[13,33,34].

## 2.2 Light Trapping with Non Planar Device Designs

Conventionally organic solar cells are created in planar configuration. Conversely, non-planar device designs can be employed to increase the photon flux within the active layer. It involves texturing the entire device, such that once light enters through the transparent window it gets completely trapped within the active region of the device. This approach can improve the device performance by increasing the short-circuit current in solar cells. It is commonly exploited in inorganic solar cells[35,36]. The biggest challenge of this approach is that in addition to the modified optical properties, the electrical properties are also affected because of the change in device architecture. Also it is difficult to extend this concept to organic solar cells because of the problems associated with the device fabrication. As organic solar cells are aimed at large scale roll-to-roll processing for achieving better economics, having non-planar device structure will be tough to implement. With this in mind, several attempts are being made to explore this approach and tailor a technique suitable for organic solar cells. Among several published reports, selected techniques are discussed here to have an overall understanding of the approach and challenges associated with them.

**2.2.1 Textured Substrate Platform**

Texturing the substrate itself and replicating the texture to the entire device built on it is one way to achieve non-planar device configurations. Since most organic based solar cells are produced by solution process, this method becomes challenging because of the difficulty in achieving smooth film over non-planar substrates. However, efforts are being taken to design techniques by studying the effect of feature size in the substrate with respect to the smoothness of the film created over them. Suitable techniques have to be developed for achieving the required feature size (mostly micro or nano scale) and fabrication of device on the textured substrates.

Nalwa et al.[37], have demonstrated this approach by using a laser interference lithography to develop 1D - diffraction gratings on the substrate. They have fabricated the device on that substrate by spin coating further layers. Using optical simulations and experiments, they report patterns with 2 μm pitch and 300 nm height on the substrate as suitable feature size for accomplishing smooth and conformal films without any defects. They have achieved ~20% enhancement in the power conversion efficiency of P3HT:PCBM devices, mainly because of increased short circuit current. The unchanged open circuit voltage and fill factor with respect to a planar controlled device shows that their device is not affected by any kind of pinholes or shunting because of the textured substrate. The broad band light trapping mostly at the band edge attributes to such enhancements. Further studies are to be made to implement this for large scale production.

Another such approach is reported by Niggemann et al.[38], where they have used a functional microprism substrate to fabricate the device. With this technique, they report the use of highly conductive polymer layer with a metal grid to replace ITO as an electrode, in addition to the advantage of light trapping effect. Acrylic polymer microprisms were made by UV casting from structures (~100 µm) obtained by precision machining. Lift-off evaporation techniques were further used to deposit metal grids. Further layers were spin coated on the microprism substrate. Even though they have demonstrated this technique with a BHJ device, the performance is not good. The increased short-circuit current because of the textured substrates gets compensated because of the loss in open-circuit voltage and fill factor due to electrical shunts in the device. This kind of problems is key in the exploration of texturing the substrate as a solution for achieving non-planar devices.

**2.2.2 Wrinkles and Deep Folds**

It is difficult to engineer techniques for placing micro or nano scale features on the substrates at specific distances for creating the textured devices. Utilization of the self-organizing responses such as wrinkles and folds is an alternative to create texture within the devices. This kind of approach is explored in devices because of its suitability for large scale production[39]. Also, devices made on these kind of structures (wrinkles and folds) are much reported to be much more robust[40].

One such approach has been demonstrated for organic solar cells by Kim et al.[39] They utilize equi-biaxial compressive stress to create wrinkles and folds on the surface of an optical adhesive film supported by a glass or plastic substrate. These wrinkles and folds localize and trap the light effectively and increases the light harvesting of the device. This has been demonstrated using a P3HT:PCBM solar cell formed over the wrinkles and folds on the substrate. A representative J-V characteristics is given in Figure 2-1. It is observed that a composite surface (both wrinkles and folds) is better than surface with only wrinkles. The increased absorption beyond the absorption edge accounts for the improved performance for devices on composite surfaces. A 47% enhancement in short-circuit current is reported for devices on composite surfaces with respect to a planar device. Using optical simulations, it is confirmed and explained that light trapping and wave guiding is the mechanism for such enhancements.

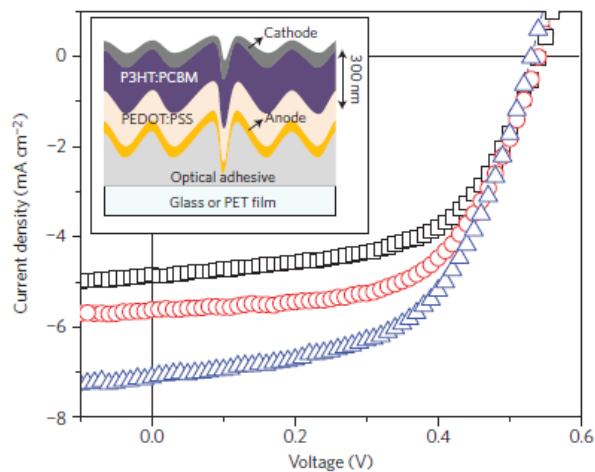

Figure 2-1. J-V characteristics for P3HT:PCBM solar cells constructed on flat (black squares), wrinkled (red circles) and composite (blue triangles) surface as demonstrated by Kim et al.[39] (Figure inset shows the device structure with wrinkles and folds)

In addition to the advantages of large scale application and robustness that can be achieved in this kind of approach, it can be used to achieve absorption beyond the absorption edge of the active layer material. It is of great potential for improving the performance of devices based on low band gap organics.

## 2.3 Optical Structures with Planar Device Design

As discussed above (Section 2.2) non-planar device designs with texture over the entire device area are hard to implement for organic solar cells. Another possibility is to add some optical structures on the surface of the planar device. These kind of structures are separated from the active layer and are applied either to the light incoupling surface or at the rear end. This approach is convenient because of ease to fabricate and integrate the optical structures in OPVs. They just change the optical properties without affecting the electrical properties. The optical structures contribute to increase the amount of light within the active region of the device resulting in an improved short-circuit current and power conversion efficiency.

There have been few reports investigating the potential of external optical structures. Some of these are discussed here to give a complete understanding and idea about the fabrication and incorporation of the optical structures in OPVs. The concept of textured rear reflector demonstrated in Chapter 3 of this thesis is one such example.

### 2.3.1 Light Trapping by Textured Films

Most organic solar cells are built in superstrate configuration with glass substrate acting as the window for solar light. Due to difference in the refractive index between glass and air, ~4% light is lost because of reflection during illumination. In this manner, considerable amount of light is reflected back at subsequent interfaces and the overall reflectance could be very high. To avoid this, anti-reflective (AR) coatings are commonly used in Si based solar cells[41]. Such AR coatings are based on dielectrics, graded index structures or textured surfaces. Among these, textured surfaces are more suitable and relevant to organic solar cells. It involves use of micron-sized polymer structures to avoid the reflection losses at the air / glass interface. Also, the textured structures used on the incoupling surface can be adjusted to facilitate light trapping.

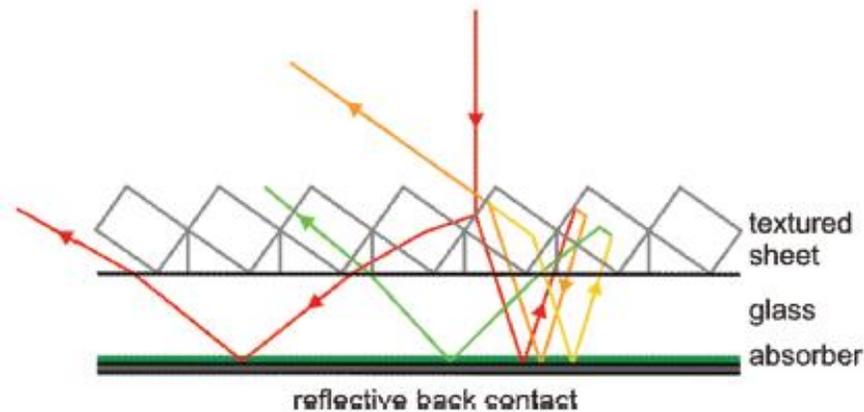

Figure 2-2.Schematic of the trajectory of light path showing the reflection at the back side of the textured sheet as illustrated by Esiner et al.[42]

One such technique has been demonstrated by Esiner et al.[42], in a polymer solar cell. They used a stainless steel mold to replicate a retroreflective textured sheet made of PDMS. The sheet is flexible and made of tilted cubic structure (in the order of 100's of micrometers) arrays, which can be applied to the substrate easily after the device fabrication. In addition to avoiding the reflection losses, the textured sheet also facilitated light trapping because of the angles at textured polymer / air interface (Figure 2-2). The rear surface of the sheet reflects back the unabsorbed light into the active layer with longer path lengths. Approximately 19% enhancement in short circuit current was achieved with the textured sheet. Similarly, Cho et al.[43], have utilized V-groove texture to achieve ~15% enhancement with tuned vertex angles for the grooves. These kind of approaches are promising because of its ease to incorporate with device without affecting the fabrication procedure.

**2.3.2 Patterned Mirror and Lens Light Trap**

Because of thin active layers used in organic solar cells, the light entering the device is not completely absorbed. Even with the use of reflective electrode at one end, the light is subjected only to two passes. After this, substantial amount of light leaves the device unabsorbed. Design of trapping elements at the light incident surface can avoid the escape of light and trap within the device. The trapping element suggested by Tvingstedt et al.[44], is an example for this. They utilize photolithography and wet etching techniques to produce a microlens combined with a reflective mirror which has self-aligned apertures. This element is incorporated at the incoupling surface, where the microlenses focus light to the openings in the reflective mirror and enters the device. After that the light gets reflected back and forth within the device because of the reflective back surface of the trapping element. This leads to an increased photon absorption in the active layer and unabsorbed photons are recycled.

The major drawback of this approach is that vary tiny openings in the mirror at incident surface causes loss of light initially by reflection. Also, they are highly dependent on angle of illumination which is not good. Fabrication techniques like etching and lithography are very expensive and thus approach is not suitable for incorporation at large scale. These drawbacks helps us to clearly understand about the important factors which are to be considered while designing an optical structure or light trapping element.

**2.3.3 Microlens Array**

Besides the usage of textures and light trapping elements for increasing the photon flux within the active layer, incorporation of a periodical array of hemispherical pattern known as microlens array (MLA) is an effective optical management technique.

It essentially alters the way light interacts with the device. The curved features on the surface bends the normal incident light ray due to refraction, leading to an increased path length within the active layer. Apart from this, the periodic lens structure helps to transmit the light incident at very large angles also and reduces the amount of light lost because of reflection at the incident surface. A combination of these provides an increased absorption with the use of microlens array leading to improved device performances. Figure 2-3(A) shows a schematic for the light behavior with and without microlens array. The work in Chapter 4 of this thesis is based on this concept and more details are discussed there.

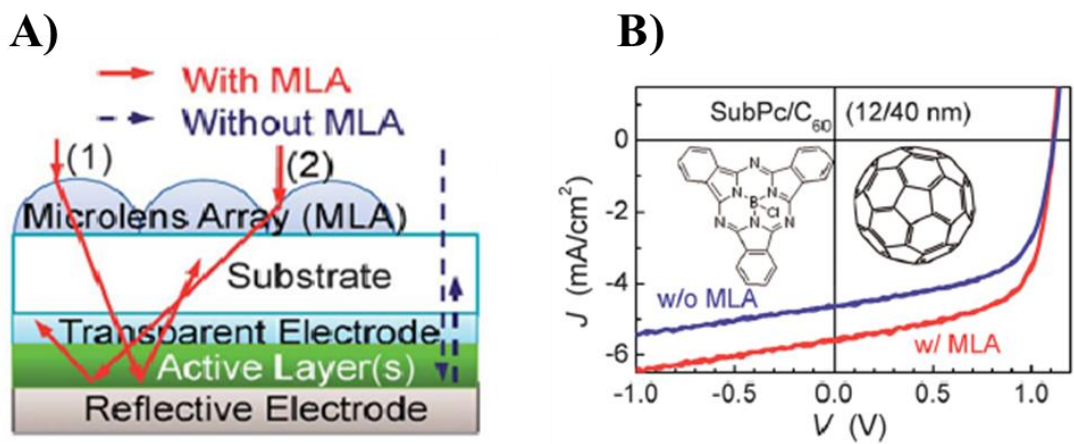

Figure 2-3. A) Schematic of the light behavior with and without microlens arrays in bilayer heterojunction device as illustrated by Myers et al.[45] B) J–V characteristics for a SubPc/$C_{60}$

Myers et al.[45], have explained this technique with a very simple fabrication procedure and it is highly promising because of the low cost and scalability. Microlens arrays are fabricated by a simple soft-lithographic technique. The structures are made with the help of a PDMS mold using Norland Optical Adhesive (a transparent polymer). They can easily be attached to the device surface externally without affecting the fabrication procedures. The lens structures are ~100μm in size. Approximately 15-60% enhancement in $J_{SC}$ is observed for a number of small molecule and polymer based organic solar cells using MLA. One example is given in Figure 2-3(B), which shows the $J$-$V$ characteristics for SubPc/$C_{60}$ bilayer heterojunction device with and without MLA. An important advantage of this technique is that the dependency on the angle of illumination is greatly reduced because of the curved nature of lenses. This approach has great potential because of its cost effectiveness and compatibility with high throughput and roll-to-roll processing of organic solar cells.

**2.3.4 Pyramidal Rear Reflector**

The approaches discussed above (Section 2.3.1 to 2.3.3) involves addition of optical structures to the light incoupling surface. However, texturing and addition of optical structures can be done at the rear reflecting end. These kind of structures facilitates the reflected light to have longer path length and achieve light trapping within the active layer. In this approach, a transparent device is required such that the textured rear reflector is attached to one end and the other side is used for illumination. The prospects of this methodology is relatively less explored for organic solar cells. Only few reports are available, which uses this kind of optical management.

Cao et al.[31], have shown the use of a pyramidal rear reflector as a light trapping design. They have used a soft lithographic technique to form pyramidal structure (with base angle $30^0$) using a transparent polymer. Rear reflector is realized upon metallization on the surface of this structure. It can be easily attached to the external surface of the device without affecting its electrical performance. The pyramidal rear reflector facilitates four passes of light through the active layer (as shown in Figure 2-4(A)), resulting in an approximately 2.5 times greater path length with respect to a planar reflector. Using the pyramidal reflector, around 11-75% enhancement in $J_{SC}$ is observed for devices of varying size and different active layer thickness. One example is shown in Figure 2-4(B), which has a comparison of J-V characteristics[31] for P3HT:PCBM based OPV device with planar and pyramidal reflectors. However, it is observed that this approach has a greater dependence on the position and angle of illumination. This could be a roadblock for the practicality of this approach.

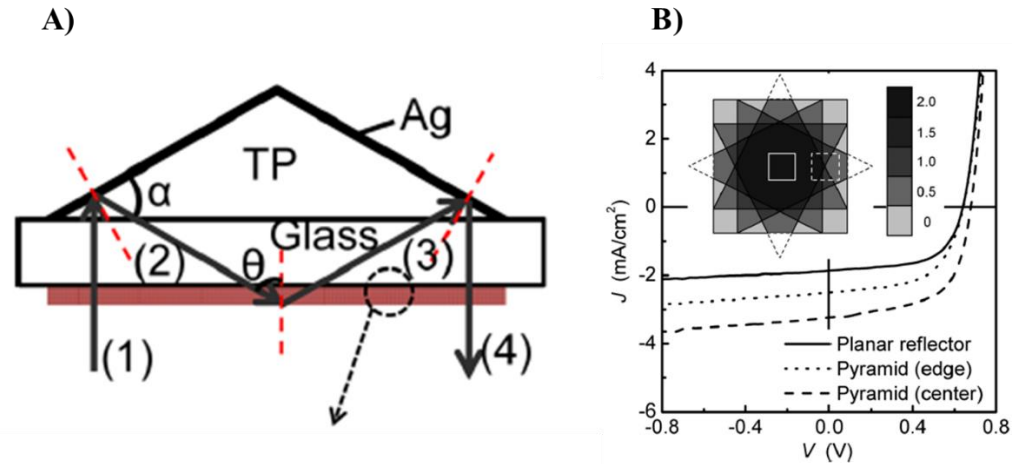

Figure 2-4. A) Schematic illustration for four passes of light through active layer with a pyramidal rear reflector by Cao et al.[31] B) J–V characteristics of P3HT:PCBM solar cell devices with planar or pyramidal reflectors as demonstrated by Cao et al.[31]

The textured rear reflector demonstrated in Chapter 3 is based on the similar concept of texturing rear end and explains further details of this approach.

## 2.4 Other Light Trapping Ideas

Apart from the usage of texturing for optical management, a number of techniques based on other ideas are being investigated and demonstrated. Some of these ideas are outlined here but not discussed in detail. Instead a brief idea is provided to understand the directions in which the research attempts are being made.

### 2.4.1 V-Shaped Light Trapping Configuration

Rim et al.[46], have demonstrated a light trapping scheme by fabricating device in a V-shaped substrate. The trapping scheme is on a much larger scale in comparison to the optical structures which are usually in micrometer and nanometer size range. The incident optical rays bounces back and forth several times through the active layer because of the V-shape and the density of bounces increases with decrease in the opening angle of the V-fold. It can be used for OPVs with active layers thickness in the order of wavelength of light or less.

### 2.4.2 Optical Fiber Solar Cells

Intrinsic light trapping designs are developed by forming organic solar cells on an optical fiber[47,48]. The device fabrication involves dipping the fiber in a polymer solution to form the active layer followed by subsequent thermal vacuum deposition of the metal. The light is coupled and trapped within the optical fiber as it passes along the axis of the fiber and makes multiple passes through the active layer. This leads to a considerable enhancement in the external quantum efficiency of organic solar cells. These kind of structures based on optical fibers does not have dependence on the angle of illumination and greatly avoids the shading effect of planar devices.

### 2.4.3 Luminescent Solar Concentrators

Luminescent solar concentrators (LSCs) can be used to achieve high level of optical concentration in organic solar cells. It involves deposition of organic dyes on a high refractive index glass substrate. These organic dyes absorb the sunlight and reemits the light at a lower energy which are trapped within a waveguide because of total internal reflection. The increased photon flux by LSCs provide great improvements in the external quantum efficiency of organic solar cells. Currie et al.[49], demonstrated this concept using a tandem organic solar cell. They use shorter wavelength dyes in a top OSC in tandem with longer wavelength dyes in a bottom OSC. By this technique, the unabsorbed light transmitted by the top OSC can be harvested by a lower band gap bottom OSC with the help of longer wavelength dyes.

### 2.4.4 Light Trapping with Nano-Structures

Conventional light trapping elements involve texturing the surface to utilize the total internal reflection and allow a longer duration for the light to stay within the active layer. However, based on ray optics the maximum absorption enhancement is limited to $4n^2$, where $n$ is the refractive index of the active layer material[50]. This limit can be surpassed by the use of periodic nano-structures and enhance the light absorption within the active layer. Corrugated metallic gratings, nano-particles embedded in the surface or the photoactive layer are some techniques used to trigger the generation of surface plasmons. The surface plasmons facilitate high optical absorption in OPVs with optically thin films. The mechanism for enhancement by surface plasmons is explained in detail elsewhere[50–52]. Incorporation of sliver nanoparticles in an anodic buffer layer[53], metal gratings in back electrode[15], nano-patterned anodes[54] are some demonstrated examples for utilization of surface plasmons in enhancing optical absorption. Photonic crystals[50] are another kind of trapping element which are not discussed here.

### 2.5 Summary

In this chapter, a review of existing optical management techniques for organic solar cells is provided. The approach of texturing the entire device to facilitate light trapping is useful, but tough to implement because of scalability and cost efficiency. The optical structures like textured films, microlens array and pyramidal rear reflectors are universal techniques and can be used with a number of material systems and device architectures. They also have the potential to be incorporated in high throughput, large-scale processing. Significant research is required to achieve practicality of other ideas like V-shaped light trapping, optical fiber, luminescent concentrators and nano-structures. Thus, optical management approach is promising and could play a vital role in achieving high efficiency OPVs.

# CHAPTER 3
# LIGHT HARVESTING IN ORGANIC SOLAR CELLS WITH A TEXTURED REAR REFLECTOR

## 3.1 Introduction

Optical Management is a promising approach to improve the efficiency of organic solar cells. This approach involves incorporation of external / internal structures to alter the way light interacts with / within the solar cells. The optical structures facilitate the enhancement of light harvesting within the absorbing layer leading to an overall increase in power conversion efficiency of organic solar cells. As discussed in Chapter 2, a limited number of techniques have been proposed for light management in OPVs.

In this chapter, a textured rear reflector is proposed as a solution for enhancing the light entrapment in organic solar cells. The concept of textured rear reflector is introduced in Section 3.2.1. The textured rear reflector causes an increase in path length of light along with trapping the light within active layer (Section 3.2.2). The design of textured reflector used in this work is explained in Section 3.2.3. Section 3.3 provides experimental details about fabrication of devices (3.3.1), microlens array (3.3.2) and incorporation of textured rear reflector with devices (3.3.3). It is observed that an optimum thickness of ZnO is to be used to achieve a balance between the leakage of light and concavity of the reflector (Section 3.4.1). The amount of light absorbed by the active layer in the first pass determines the effectiveness of a textured reflector (Section 3.4.2). Some general guidelines about the suitability for enhancement using textured reflector are outlined in Section 3.4.3. Finally, the results of this chapter are summarized in Section 3.5.

## 3.2 Textured Rear Reflector

### 3.2.1 Concept

Conventionally OPV device designs are planar, which incorporates a transparent electrode on one side and reflective electrode on the other side. They may be used in either substrate or superstrate configurations. In such designs, for normal incidence the light passes twice through the active layer because of the reflective electrode (Figure 3-1). The path length of light in the active layer is limited to $2t$, where $t$ is the thickness of active layer. At the end of second pass, light unabsorbed by the active layer is lost permanently. This sets a limit to light absorption and in turn device efficiency of OPVs.

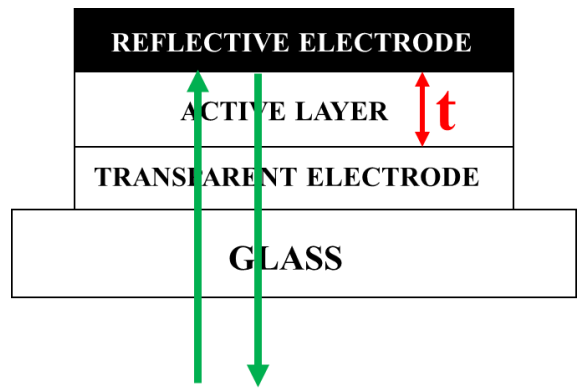

Figure 3-1. Light path in a conventional OPV device with an active layer of thickness, $t$.

A number of studies have been made to understand the effect of texturing the light incoupling surface of OPVs in order to increase the path length of light (Section 2.3). Enhancements up to 60% have been demonstrated and the results are promising[45]. Texturing the reflector at the rear end of device is another useful approach to increase the path length of light absorbed within the active layer (beyond $2t$). The mechanism for enhancement is similar to that of texturing the incoupling surface. However, there have been very few studies related to texturing of the rear end of device (Chapter 2).

### 3.2.2 Mechanism for Enhancement

The feature size of the textured reflector used in this work is in the order of few microns (~2 – 4 µm). This is larger than the wavelength of visible light (~390 – 700 nm) and thus comes under the "ray optics regime". The concept of surface plasmons is not applicable here as it requires nano-sized features. Figure 3-2(A) and (B) shows a schematic of the light path for normal incidence in semi-transparent device structures with planar and textured reflector respectively at the rear end. The requirement to use a semi-transparent device is explained further in Section 3.2.3. It can be observed from Figure 3-2(A) that with a planar reflector, the path length of light absorbed within the active layer during each pass is $t$ and the unabsorbed light after the second pass is lost from the device. The total path length is $2t$ and this situation is same as a conventional OPV device structure (Figure 3-1). In the device with textured reflector (Figure 3-2(B)), the path length of light during the first pass is $t$ (same as planar reflector). However as the light reaches the textured surface of the reflector, it has a non-normal incidence as opposed to normal incidence in a planar reflector. Let $\theta_i$ be the angle between the incident ray of light and normal at the point of contact in the textured surface. Assuming a specular reflection, the angle of refection $\theta_r$ is same as the angle of incidence $\theta_i$ (i.e.) $\theta_i = \theta_r$. The reflected light ray is at an angle $\theta$ from the device normal. From the Figure 3-2(B), the relation between $\theta_i$, $\theta_r$ and $\theta$ can be given as $\theta = \theta_i + \theta_r = 2\theta_i$. The path length of reflected light thus has an additional angular component of $1/\cos\theta$. Consequently the total path length of light in a textured reflector is $t + t/\cos\theta$, which is greater than $2t$ in a planar reflector.

In addition to this, the textured reflector also has the potential to reflect back the unabsorbed light. As shown in Figure 3-2(A) and (B), the light unabsorbed after the second pass reaches the glass substrate / air interface. Considering the refractive index of air ($n_{air}$) and glass substrate ($n_{glass}$) as 1 and 1.5 respectively, the critical angle ($\theta_C$) to facilitate the total internal reflection at glass / air interface can be calculated. Based on Snell's law, $sin\theta_C = n_{air}/n_{glass} = 1/1.5 \ \ or \ \ \theta_C = 42^0$. The angle between the light ray and the device normal at the glass / air interface should be greater than $42^0$ for total internal reflection to occur. As the textured reflector reflects the light at relatively larger angles, most of the light rays reaching the glass / air interface have angles greater than $42^0$ resulting in reflecting back the light into the devices. This adds to the path length of light absorbed within the active layer. In this way, textured reflector facilitates trapping of light within the active layer. However, this is not applicable for planar reflectors.

Therefore, the above mentioned mechanisms increases the path length of light and traps the light within the active layer for devices with textured reflector. This would considerably increase the amount of light absorbed resulting in a higher short circuit current and power conversion efficiency in comparison to a planar reflector. This technique is universally applicable to thin film OPVs with any kind of active layer material and device structures.

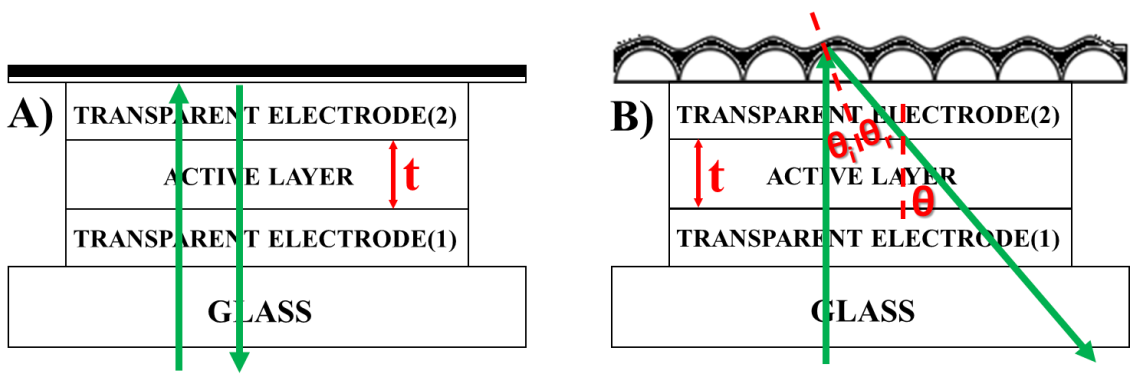

Figure 3-2. Path of light in a semi-transparent device with an active layer of thickness, *t* and reflector at one end. A) Planar Reflector. B) Textured Reflector.

**3.2.3 Design of Textured Rear Reflector**

In general, a textured surface is achieved using a number of techniques like lithography[37,55], etching[56], evaporation techniques[38], application of mechanical stress[39] etc. These techniques are not only expensive but also difficult for adopting at a larger scale. In this work, a relatively simpler approach is developed to fabricate and incorporate a textured reflector which has a potential for large scale production. A unique soft lithography technique have been developed by our previous group members[57] to fabricate microlens arrays (MLA). They have demonstrated the incorporation of MLA to the incoupling surface as a universal approach to achieve light entrapment in OPV devices[45]. The fabrication procedure for MLA based on the soft lithography technique is detailed in Section 3.3.2. In this work, the microlens array is used to design a textured reflecting surface. The microlens array is made of Norland Optical Adhesive (NOA), which is a transparent polymer (Section 3.3.2). A reflective metal can be deposited on the transparent MLA to form a reflective surface, with the texture of the MLA surface reciprocated. The resulting textured reflective surface would be concave on one side and convex on the other side (Figure 3-3).

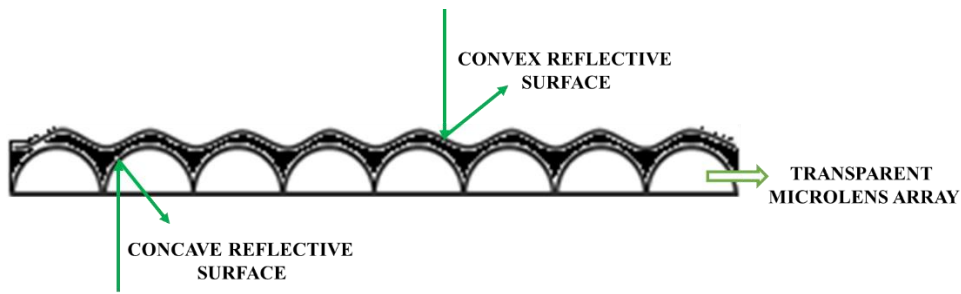

Figure 3-3. Schematic of a textured reflector showing different reflective surfaces.

However, it cannot be done simply in such a manner. Thermal vacuum evaporation is the most commonly used method for metal deposition in OPV fabrication and it involves placing the source material at a certain distance from the substrate for deposition (Section -1.7.1). The height of periodic structures in the surface of microlens array causes a shadowing effect during the thermal deposition of metal. As a result of this effect, the metal film does not completely cover the entire area of MLA. As shown in Figure 3-12(A) and 3-13(A), there is a poor or no coverage in the deeper portions of the MLA surface. Intuitively, it is understood that the shadowing effect cannot be eliminated irrespective of the thickness of metal film deposited.

In this work, nanoparticles are used as a way to minimize the shadowing effect and improve the coverage of the metal film. The particles are small enough that they can cover the deeper pits in the MLA surface and make the metal film conformal. The important requirements for the filling materials are high transparency and a refractive index lower than MLA. With a low refractive index than MLA, internal reflection becomes possible at MLA / ZnO interface. In addition to this, having a high transparency helps to avoid any optical losses because of the filling material. Zinc oxide (ZnO) nanoparticles used here have high transparency due to its large band gap. However it doesn't have a lower refractive index than MLA. Regardless of this, ZnO was used because of material limitations. One solution to overcome this is using MLA of higher refractive index than the filling material used (Section 5.2). Further, as discussed in Section 3.4.1, an optimum thickness of ZnO has to be used to achieve a balance between the leakage of light and texture of the reflector surface. Additional details about the fabrication and incorporation of textured reflector in devices are given in Section 3.3.3.

Therefore, as elucidated above a relatively simpler approach has been used to form a textured reflector using a combination of MLA, ZnO nanoparticles and metal film.

## 3.3 Experimental Details

### 3.3.1 Device Fabrication

The conventional solar cell design (Figure 3-1) with one transparent and one reflective electrode cannot be used for incorporation of a textured rear reflector. The device has to be developed with two transparent electrodes, such that the textured reflector can be incorporated on one side and the other side can be used as a window for sunlight to enter the solar cell. As discussed in Section 1.6 the transparency of the electrodes play a critical role in the ultimate device efficiency achieved. Indium Tin Oxide (ITO) and Oxide/Metal/Oxide (OMO) trilayer structure are the choice of two transparent electrodes used in this work. The transparency of the OMO trilayer used in this work is lower in comparison to ITO (Figure 3-4) and the devices are thus semi-transparent. In spite of the reduced transparency, semi-transparent devices are used here because the ultimate purpose is to study the effect of textured reflector in comparison to a planar reflector. Better choice of electrodes can be used to achieve higher overall efficiencies.

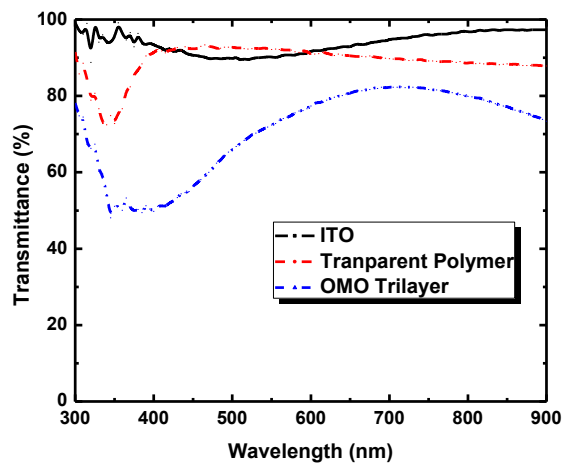

Figure 3-4. Transmittance of ITO, Transparent Polymer (Norland Optical Adhesive) and OMO Trilayer (10nm $MoO_x$/15nm Au/50nm $MoO_x$).

The devices are fabricated on ITO patterned glass substrates. The substrates are rigorously cleaned to remove all the impurities. The cleaning procedure involves sonication in DI Water + Soap, DI Water, Acetone and Isopropyl Alcohol (IPA) for 15 minutes each. Finally just before starting the device fabrication, the substrates are subjected to 15 minutes of UV ozone treatment, during which ultraviolet light along with ozone cleans and modifies the work function of ITO surface. The cleaned substrates are used to fabricate a device structure as shown in Figure 3-5. A 40nm zinc oxide film is formed by spin coating of ZnO nanoparticles suspended in ethanol solution, which is prepared using a sol-gel method. Then, the ZnO film is annealed at $70^0C$ in a $N_2$ filled glovebox for 15 minutes to facilitate evaporation of solvent.

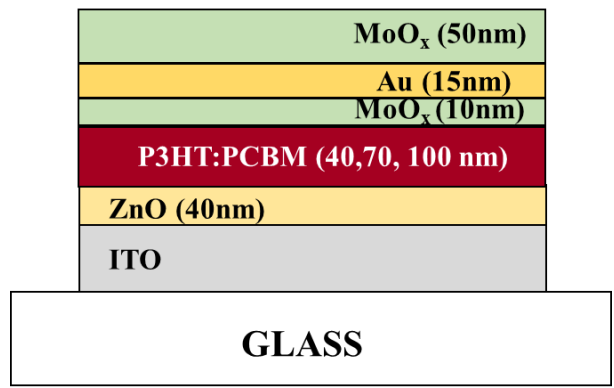

Figure 3-5. Structure of a semi-transparent device with two transparent electrodes.

After ZnO, P3HT:PCBM solution (with 1:0.8 weight ratio in chlorobenzene) is spin coated to form the active layer. The concentration of the solution is varied to control the active layer thickness, i.e., 27 mg/ml, 18 mg/ml and 9 mg/ml are used to form 100nm, 70nm and 40nm active layer respectively. Then, the P3HT:PCBM film is annealed at $150^0C$ for 30 minutes in a $N_2$ filled glovebox for the microphase separation in the blend film and to form channels for the carrier to reach the electrode. After the annealing, OMO trilayer consisting of 15nm Au sandwiched between 10nm and 50nm of $MoO_x$ is deposited using thermal vacuum deposition to form the second electrode and complete the device. The area of devices used here is 2x2 mm$^2$.

### 3.3.2 Fabrication of Microlens Array

The formation of microlens array (MLA) is the first step in the fabrication of a textured rear reflector. Figure 3-6 schematically shows the steps in the fabrication of MLA. The template for MLA fabrication is formed using Langmuir-Blodgett (LB) technique. A self-assembled SiO$_2$ monolayer on a glass substrate is formed using this technique. First SiO$_2$ particles with hydrophobic ligands (4μm size) were dispersed in 1,2-ethanediol. The dispersed solution is dropped on the surface of water in a container, with a piece of cleaned glass substrate already immersed in it. The hydrophobic ligands causes the SiO$_2$ sphere to immediately flow on the surface of water. As the glass substrate is slowly pulled out from the water in the container (Figure 3-6(A)), a well assembled monolayer of the spherical SiO$_2$ particles are formed on the glass substrate due to capillary effect. The monolayer of SiO$_2$ spheres is closely packed with very few defects, as shown in optical micrograph of the template formed on glass in Figure 3-7(A). One of the defect region is magnified in the figure inset. The SiO$_2$ spheres are ~4μm size and the final microlens array will be of same size.

Then, poly(dimethylsiloxane) (PDMS) precursors with a 10:1 weight ratio of polymer to curing agent are mixed and kept in vacuum to drive away the air bubbles. Once it is free of bubbles, the PDMS precursors are poured onto the micro-lens template on glass and cured at 60$^0$C in a vacuum oven for 1 hour. After it is completely cured, the PDMS is slowly peeled off from the glass template (Figure 3-6(B)). The resulting PDMS mold has concave lens features which is in order of 4μm, same as the size of initial SiO$_2$ spheres used.

The microlens array is formed on the device or glass surface using Norland Optical Adhesive (NOA), which is a transparent polymer. It has a transparency (>90%) in the visible wavelength range, as shown in Figure 3-4. A drop of NOA is placed at the center of the substrate onto which MLA has to be formed. The mold is then gently placed on the surface and held for 3 minutes to allow the NOA to fill the concave gaps in PDMS mold and spread over the entire substrate. It should be made sure that the mold doesn't move and damage the device contacts. The NOA is then cured using a 365nm ultraviolet light.

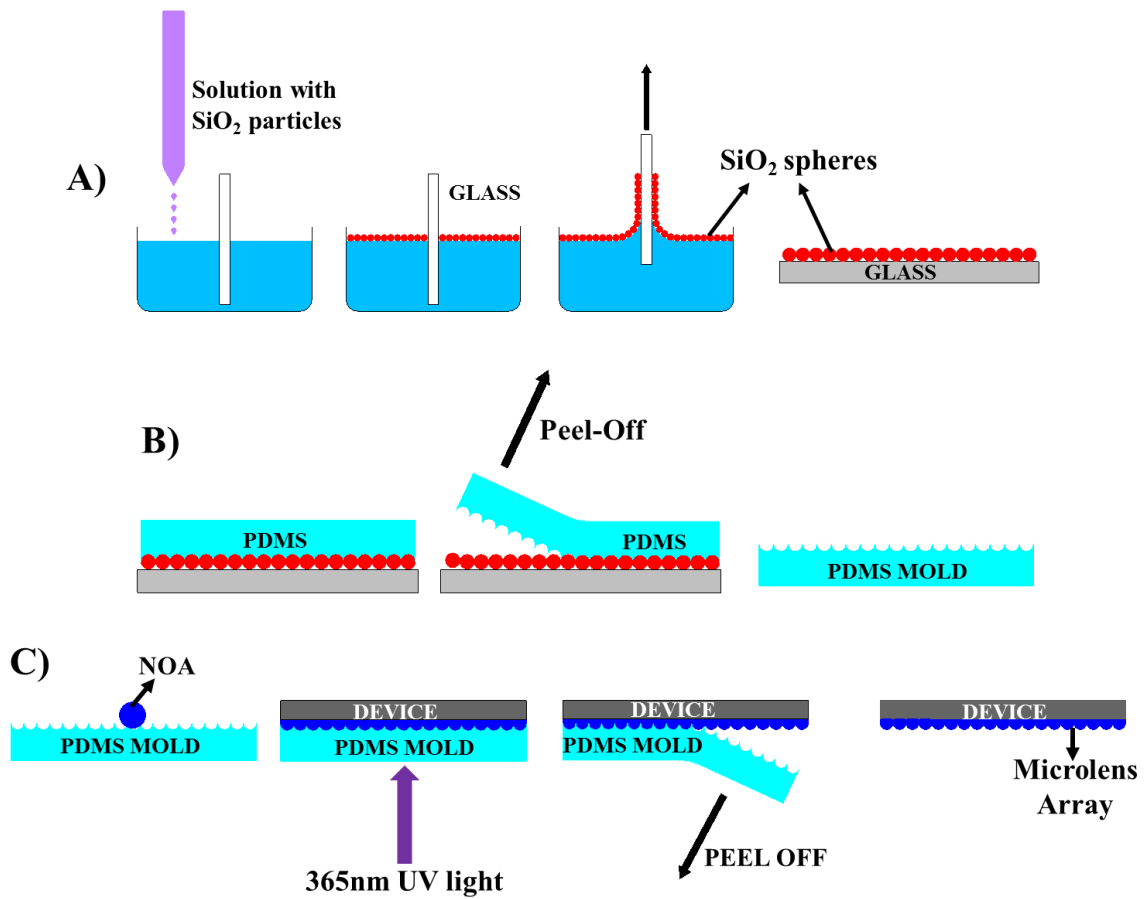

Figure 3-6. Schematic for fabrication of microlens array (MLA). A) Steps in Langmuir-Blodgett method for microlens array template fabrication. B) Formation of PDMS mold using peel-off technique. C) Steps in forming MLA on the device using a transparent polymer, Norland Optical Adhesive (NOA).

The time for curing depends whether MLA is formed on the glass substrate or on the transparent OMO trilayer electrode. The required time differs due to the difference in a number factors like amount of light absorbed by the underlying layers, surface energy of the adhering surface etc., It is observed that a curing time of 7 minutes and 12 minutes is required for forming on glass and trilayer respectively.

After the curing is complete, the PDMS mold is slowly peeled off from the device forming a microlens array of the surface. The PDMS mold can be used several times to form the MLA and eventually it may get damaged due to its exposure to ultraviolet light and repeated usage. Unless mentioned, 4µm sized microlens arrays are used in this work. Figure 3-7(B) shows the optical micrograph of 4 µm sized microlens arrays with some defects. A scanning electron microscope (SEM) image of 4 µm MLA is given in the figure inset. The defects in MLA are replicated from the PDMS mold.

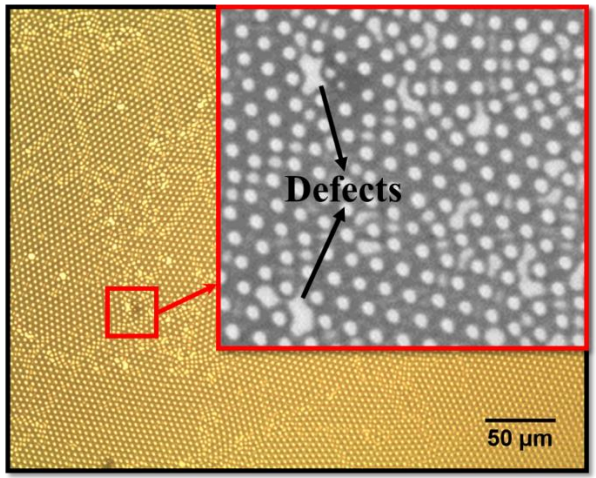
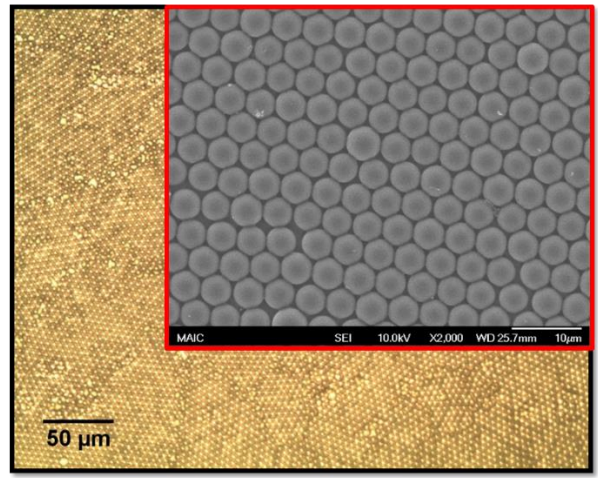

Figure 3-7. Optical microscope Images. A) Microlens array template (4µm $SiO_2$ spheres on glass) with some discontinuities (Defects are shown in the figure inset, which is a magnified image of the marked region). B) 4µm microlens array (Figure inset shows a SEM image of the MLA).

### 3.3.3 Incorporation of Textured and Planar Reflector

The planar and textured reflectors were incorporated on the OMO trilayer electrode rather than the glass side, as shown in Figure 3-2. If the textured reflector is incorporated on the glass side, a thick glass is present between the device and the textured reflector. The glass interface has an adverse effect on the interference within the device, resulting in less optical absorption by the active layer (see Section 4.4.2 for more details). On the other hand with the textured reflector on the OMO trilayer, the glass interface is avoided and the textured reflector is in direct contact with the device. This could result in a strong interference within the device leading to better optical absorption. Further advantages of incorporation on OMO trilayer is discussed in Section 4.2.2.

In addition to this, the incoupling of incoming light in the device is much more effective through glass side in comparison to the OMO trilayer side. From Fresnel's equation for normal incidence of light, we know that the amount of light reflected at any interface is given by $R = |(n_1 - n_2)/(n_1 + n_2)|^2$, where $n_1$ and $n_2$ are the refractive indices of the first and second medium respectively. Considering the refractive indices for air, glass and trilayer electrode as 1, 1.5 and 2.0 respectively, the fraction of incoming light which is reflected can be determined. It can be calculated that ~4% of light is reflected at the air/glass interface, whereas ~11% light is reflected at air/trilayer interface. This clearly demonstrates that light incoupling is better through glass side. Thus, better incoupling of incoming light along with effective light trapping makes trilayer side as the best choice for textured reflector incorporation.

Figure 3-8 shows the schematics for steps in fabrication of textured reflector. As a first step in the fabrication of a textured rear reflector, a microlens array is formed on the OMO trilayer (Figure 3-8(A)) using the procedure described in Section 3.3.2. The reflective metal film cannot be directly deposited over the microlens array because of the shadowing effect (Section 3.2.2). This scenario is explained further in Figure 3-12 and discussed in section 3.4.1. Zinc oxide nanoparticles suspended in ethanol solution are spin coated over the microlens array to fill the deeper pits and thus avoid the shadowing effect (Figure 3-8(B)). After spin coating, the film has to be annealed. Thermal annealing for ZnO mentioned in Section 3.3.1 cannot be used to anneal the ZnO on MLA, because some cracks are observed in the MLA film as shown in the optical micrograph in Figure 3-9. The cracks are due to the poor thermal stability of the MLA film and are observed to be increasing with annealing temperature. Figure 3-9(A) and (B) shows the MLA film with ZnO after annealing at $75^0C$ and $150^0C$ respectively. Hence, the ZnO films were annealed in vacuum for 30 minutes and no cracks were observed. Based on the analysis provided in Section 3.4.1, ~240-260nm is found as the optimum ZnO thickness for forming an effective textured reflector. However ~300nm is used here due to the difficulties in precise control of thickness.

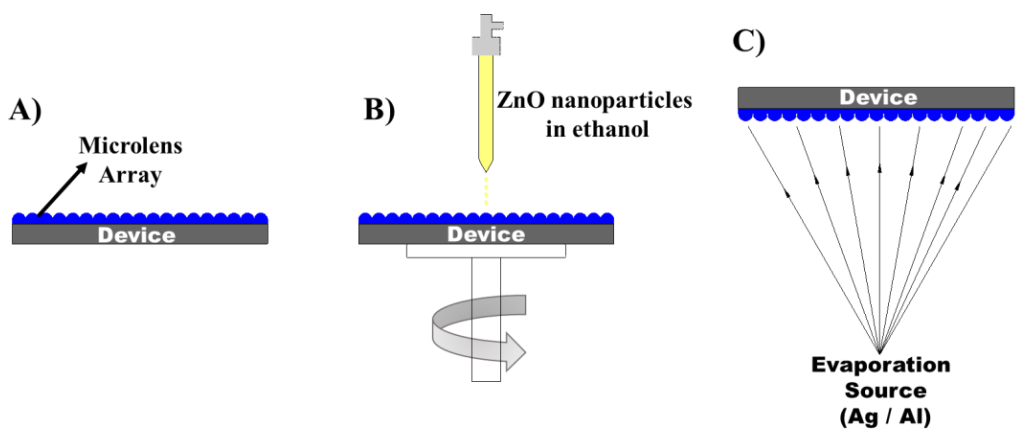

Figure 3-8. Steps in the fabrication of textured reflector. (A) Device with microlens array on OMO trilayer. (B) Spin coating of ZnO over MLA. (C) Thermal vacuum deposition of metal (Ag / Al) on MLA with ZnO.

The thickness of ZnO film can be controlled by solution concentration. It has been calibrated using profilometry that at a specific spin speed (3000 rpm/s), 30 mg/ml solution provides a 40nm film. Instead of calibrating for several solution concentration, it can be roughly approximated that solution concentration varies linearly with the film thickness at a constant spin speed, i.e. 60 mg/ml solution would provide an 80nm film and so on. This is a good enough approximation, as the precise determination of ZnO thickness is not critical. However, ~300nm films cannot be achieved by spin coating a higher concentrations of ZnO solution on MLA, as the effective spinning of solution is affected because of its high loading. It is found that 80-100 mg/ml is the highest concentration range for the solution to form considerably good ZnO films on MLA. As varying the spin speed provides only a very narrow range of thickness control, the required ZnO thickness is achieved by spin coating several layers on top of each other. A ~100nm film is formed during each step by using a solution of 80 mg/ml concentration. Vacuum annealing of 30 minutes is done after spin coating each layer. Thus, a thickness range of 0-1000nm is achieved depending on the number of layers spin coated. However, we should be aware of the error associated with the ZnO thickness.

The final step in completion of the textured reflector is thermal vacuum deposition of the metal film (Figure 3-8(C)). The metal film used here consists of 300nm silver. Silver is used here because of its better reflectivity in comparison to aluminum (which is commonly used). Thus, a textured rear reflector is incorporated on the OMO trilayer with a relatively simpler approach.

Careful incorporation of planar reflector is essential to have a good comparison and study the enhancement characteristics of a textured rear reflector. For planar reflector, a NOA blank is formed on the OMO trilayer using a blank PDMS. Then, a 300nm Ag is deposited. The NOA blank used between trilayer electrode and metal film serves dual purposes. It protects the underlying device and contacts from damage during the metal deposition. Also, as the refractive indices of NOA blank and MLA are same (n~1.58), the comparison between planar and textured reflector is valid and reliable. This is because interfaces within the devices becomes the same for both cases and the only difference between them is the texture of the reflector.

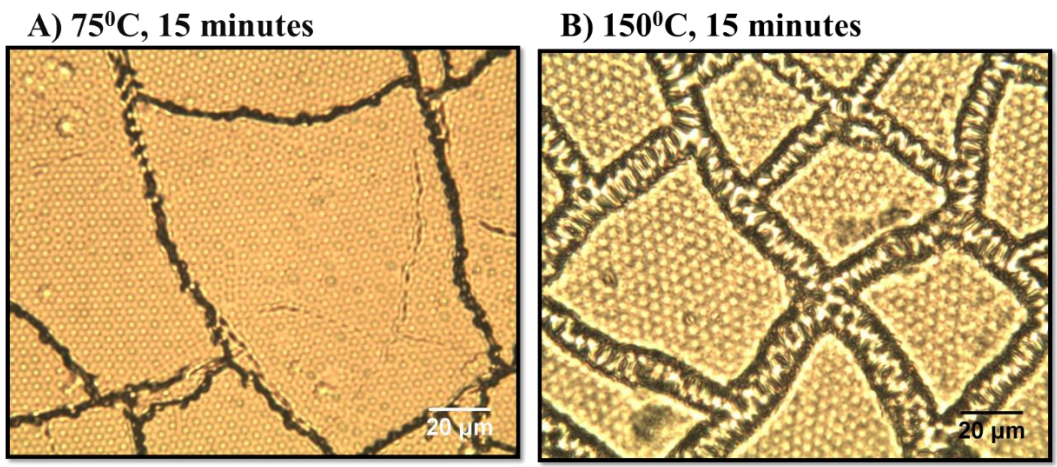

Figure 3-9. Optical micrograph of MLA film with ZnO after annealing for 15 minutes. A) $75^0$C. B) $150^0$C. Cracks are observed in the film and increases in size at higher temperature.

The area of the device is ~0.04 cm$^2$ and four devices are there in each substrate. The base area of the textured / planar reflector is ~4 cm$^2$ and is positioned at the center of the substrate without damaging the contacts. The textured reflector is made much larger than the device in order to compensate the edge effect by trapping the light which escapes around the edges of the device area. This simulates a condition similar to that of a large area device. The reflection of light at large angle by the textured reflector allows the light outside device area to be coupled within the device. The planar reflector is also ~4 cm² and positioned at the center of substrate to avoid discrepancies in the comparison

## 3.4 Results and Discussions

### 3.4.1 Optimization of ZnO Thickness in Textured Reflector

Formation of MLA on OMO trilayer is the first step in the fabrication of textured reflector. Figure 3-7(B) shows optical micrograph and SEM image of a 4 µm MLA film. The film has uniform periodic structures with just some minor discontinuities. The contact angle is important to understand the height of periodic structures in MLA film. For a 4µm lens, the height at different contact angles are calculated using diagrams as shown in Figure 3-10.

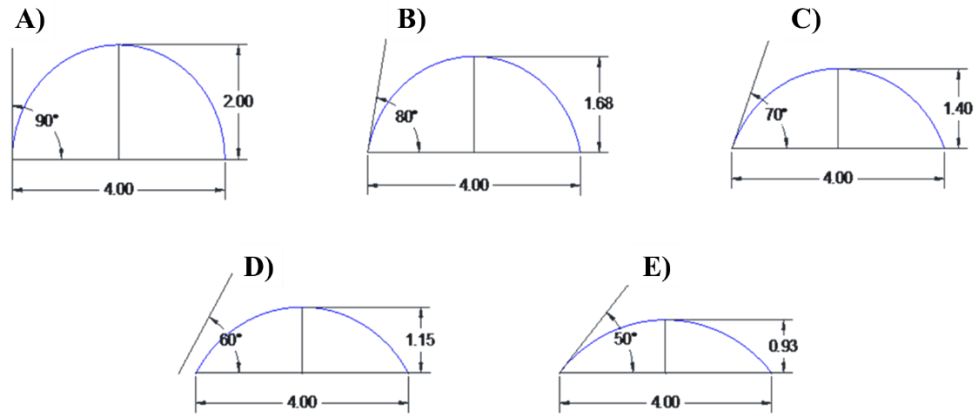

Figure 3-10. Schematics for height of 4µm lens at various contact angles. A) 90⁰. B) 80⁰. C) 70⁰. D) 60⁰. E) 50⁰.

From the Figure 3-10(A)-(E), it is observed that the height of lens decreases with decreasing the contact angle. The calculated heights are 2.00 µm, 1.68 µm, 1.40 µm, 1.15 µm and 0.93 µm for contact angles 90⁰, 80⁰, 70⁰, 60⁰ and 50⁰ respectively. Using the SEM image (Figure 3-11), contact angle for the microlens array used in this work is measured as ~$70^0$. Based on the calculations in Figure 3-9, it is expected that the height of microlens should be in the range of 1.20 -1.40 µm. The thickness profile of the MLA film are obtained through AFM measurements to confirm the microlens height.

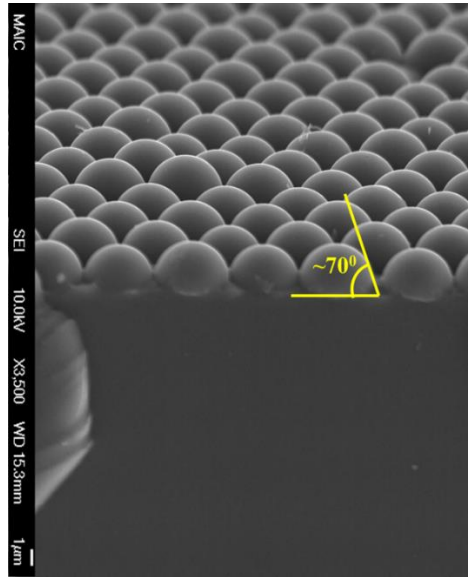

Figure 3-11. SEM image of a 4 µm MLA film showing the contact angle measurement, which is ~$70^0$ (Image Courtesy: Weiran Cao).

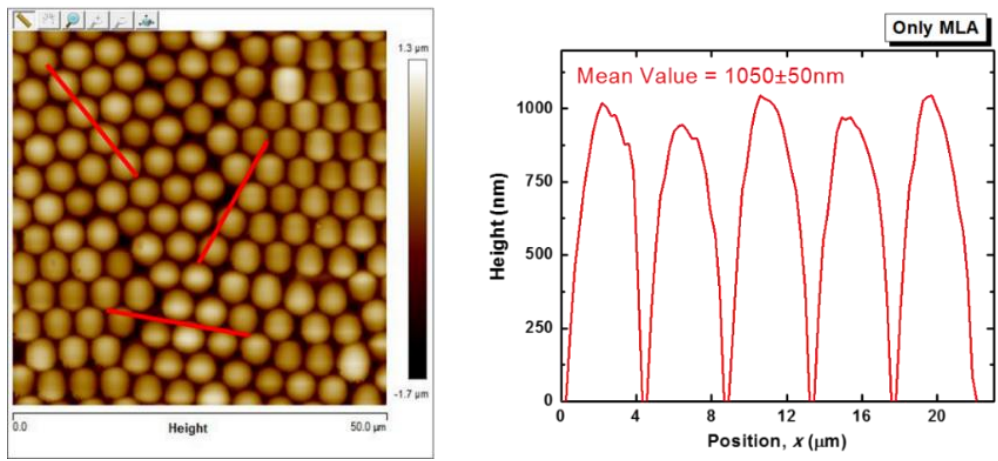

Figure 3-12. AFM image (left) of a 4µm MLA along with height analysis (right). The height is calculated using lines as shown in the left image. One such measurement is given in the right and the average height is 1.05±0.05µm.

Figure 3-12 shows the AFM image of a 4 µm MLA film. From this image, the height is calculated by drawing lines in several directions along the microlens arrays. This is not a precise technique since the lines may not exactly pass through the center (i.e.) maximum and minimum point. However, a rough estimation is obtained by averaging the values from approximately 50 microlenses. The measured value is 1.05±0.05 µm (Figure 3-12), which is close enough to the expected value (1.20 -1.40 µm). Several factors like irregularities in the mold, measurement and calculation errors can account for the difference between the values.

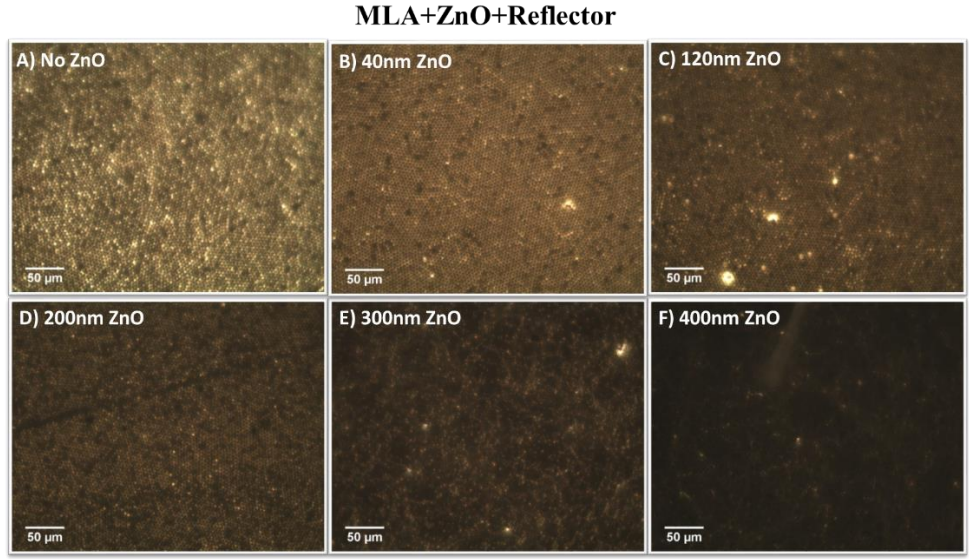

Figure 3-13. Optical microscope images (in transmission mode, fixed contrast) of textured reflector with varying ZnO thickness (Glass/MLA/ZnO/Reflector). Leakage of light is observed as tiny spots in the images. A) No ZnO. B) 40 nm ZnO. C) 120 nm ZnO. D) 200 nm ZnO. E) 300 nm ZnO. F) 400nm ZnO.

These micron sized features results in shadowing effect during thermal vacuum deposition of metal film. This makes the film coverage poor especially at deeper portions of the film (joints between microlenses) and thus the film is leaky. Figure 3-13(A) shows an optical image (measured in transmission mode) of the metal film deposited over MLA. It is observed that considerable light passes through numerous tiny spots present across the entire metal film. This correspond to the poor coverage area as mentioned above. Hence, it is confirmed that significant amount of light is lost and the reflector is not effective.

The use of zinc oxide (ZnO) nanoparticles is a good solution to get over the shadowing effect issue. The nanoparticles which are few tens of nanometers can fill the deeper pits in MLA film and provide better coverage of the metal film. Using too less ZnO may not solve the issue completely and still the film may be leaky (Figure 3-14(A)). On the other hand, too much use of ZnO can eventually flatten the MLA surface and the resulting reflector will have less or no texture (Figure 3-14(C)). Then high angle reflection, which is one of the main mechanism for enhancement through textured reflector gets limited. Thus, an optimum ZnO thickness is determined to have a balance between the two factors and develop an effective textured reflector.

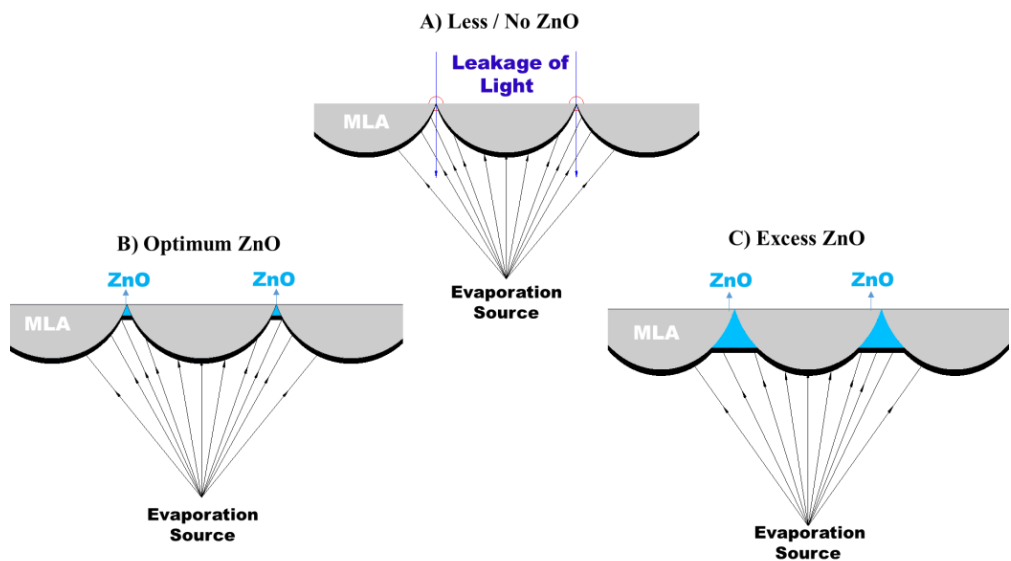

Figure 3-14. Incorporation of ZnO during fabrication of textured reflector to minimize shaddowing effect. A) Leakage of light due to less / no ZnO. B) Optimum ZnO which balances leakage of light and concavity of reflector. C) Loss of concavity due to excess ZnO.

To study this, ZnO of varying thickness ranging from 0-1000μm are spin coated over a MLA fabricated on glass. Then, the metal film is deposited to form the reflector. Optical micrographs of reflectors with varying ZnO thickness are taken in transmission mode (with contrast locked) to have a qualitative comparison and for understanding of the amount of light lost. Figure 3-13(A) - (F) shows the optical micrographs of reflectors with 0nm, 40nm, 120nm, 200nm, 300nm, 400nm ZnO thickness. It is observed that from 0nm to 400nm ZnO thickness, the images clearly show a decrease in the amount of light transmitted and gradually becomes darker. This indicates that as ZnO thickness increases, the shadowing effect gets minimized and better metal film coverage is achieved.

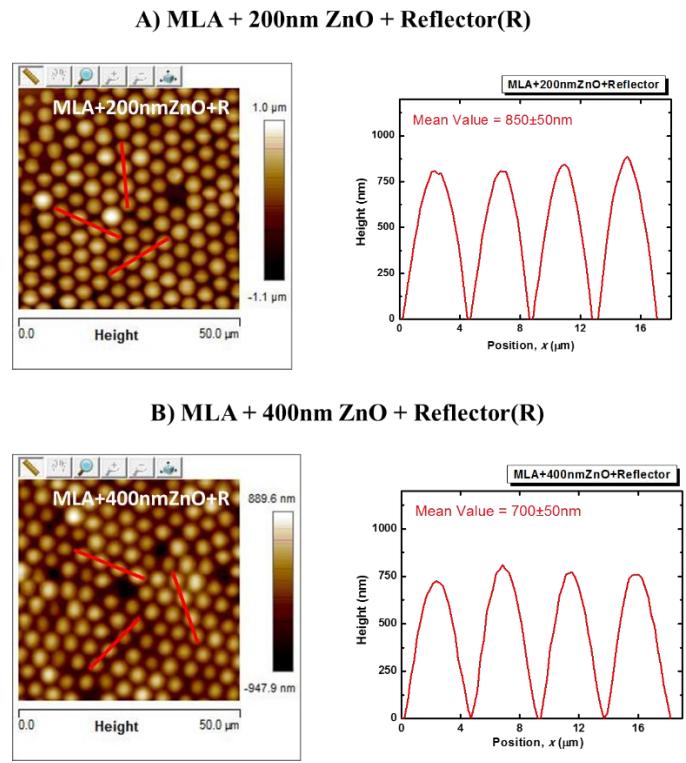

Figure 3-15. AFM image of a textured reflector (left) with different ZnO thickness along with height analysis (right). The height is calculated using lines as shown in the left image. One such measurement is given in the right image for each case. A) 200 nm ZnO (Average height = 0.85±0.05μm). B) 400 nm ZnO (Average height = 0.70±0.05μm).

For 400nm ZnO, we can still observe some faint light getting transmitted. This can be because of the minimal shadowing effect and discontinuities of the MLA layer. AFM images of textured reflectors with 200nm and 400nm ZnO are taken to verify the understanding. As the average height determined using the AFM image (by same technique described above) is the distance between the lowest and peak point, it is expected that with ZnO it should decrease appropriately. As shown in Figure 3-15(A) & (B), the height is measured as 0.85±0.05 µm and 0.70±0.05 µm for 200 nm and 400 nm ZnO respectively. These values agree with the expected values (0.85 µm and 0.65 µm) and are within the error range.

**MLA+ZnO+Reflector**

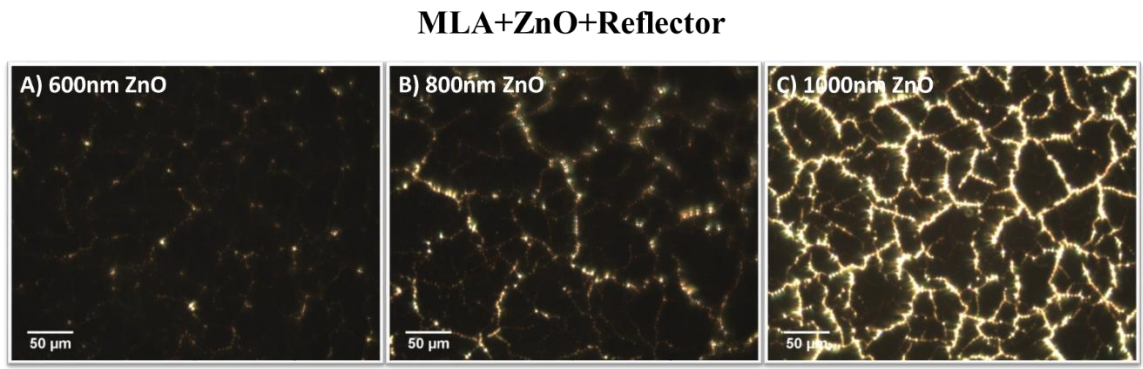

Figure 3-16. Optical microscope images (in transmission mode, fixed contrast) of textured reflector with higher ZnO thickness (Glass/MLA/ZnO/Reflector). Leakage of light is observed in the images because of the cracks in ZnO film. A) 600 nm ZnO. B) 800 nm ZnO. C) 1000 nm ZnO.

Figure 3-16(A), (B), (C) shows the optical image for textured reflector with 600 nm, 800 nm and 1000 nm ZnO respectively. It can be expected that after a certain ZnO thickness no light would be transmitted and the film will be completely dark. However for higher ZnO thickness around 600-1000nm, even if the shadowing effect is avoided completely it is observed that the amount of light transmitted increases because of the cracks in the film. The cracks grow in size with increasing ZnO. The cracks occur in the ZnO layer, which has poor stability as the thickness increases. Thus, based on the above analysis it can be inferred that ~300 – 400 nm ZnO is helpful for minimizing the shadowing effect and leakage of light along with good film stability.

On the other hand, as mentioned before excess of ZnO results in the loss of texture (concavity). To study this effect, semi-transparent devices with 40nm P3HT:PCBM active layer are made and textured reflectors with varying ZnO thickness are incorporated (Figure 3-17(C)). A semi-transparent device with (Figure 3-17(B)) and without a planar reflector (Figure 3-17(A)) are made for reference / baseline.

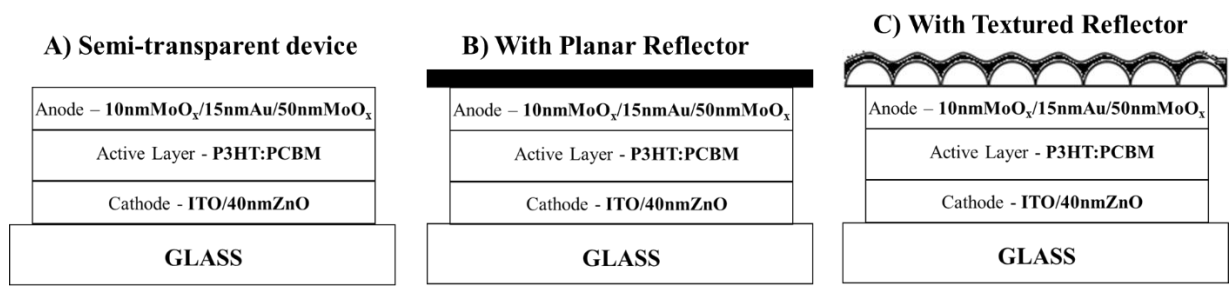

Figure 3-17. Schematics of device structures used in incorporation of a planar and textured reflector. A 40 nm P3HT:PCBM active layer was used. A) Semi-transparent device with ITO and OMO trilayer as transparent electrodes. B) Planar reflector incorporated over OMO trilayer. C) Textured reflector incorporated over OMO trilayer.

Figure 3-18 shows the short-circuit density values of the devices with varying ZnO thickness over MLA just before the metal deposition, measured through glass side. The baseline semi-transparent device ($J_{SC}$ = 2.01±0.1 mA/cm$^2$) is shown by the dotted line at the bottom. With a planar NOA blank on the OMO trilayer, no considerable change is observed and is constant ($J_{SC}$ = 2.03±0.1 mA/cm$^2$) irrespective of ZnO thickness. This is understandable because the rear surface is planar and thus there is no change in the path length of light or any kind of light trapping.

Whereas with MLA at the rear end, the $J_{SC}$ value increases from 2.01±0.1 mA/cm$^2$ to 2.51±0.1 mA/cm$^2$ due to total internal reflection at the MLA / air interface. With increasing the ZnO thickness from 0nm to 400nm, the $J_{SC}$ value decreases from 2.51±0.1 mA/cm$^2$ to 2.13±0.1 mA/cm$^2$ (~15% decrease). The refractive index (n) of ZnO nanoparticles is ~2. However, the refractive index reduces to around ~1.6-1.7 when used in films depending on the thickness and porosity. Accordingly, no reflection is possible from the MLA (n = 1.58) interfaces covered with ZnO. This leads to loss in short-circuit current. Therefore, as more MLA regions are being covered with increasing the ZnO thickness, amount of large angle reflection back into the devices gets limited. This explains the decrease in $J_{SC}$ value with increasing ZnO. Therefore, it is evident that excess ZnO leads to loss of texture (concavity).

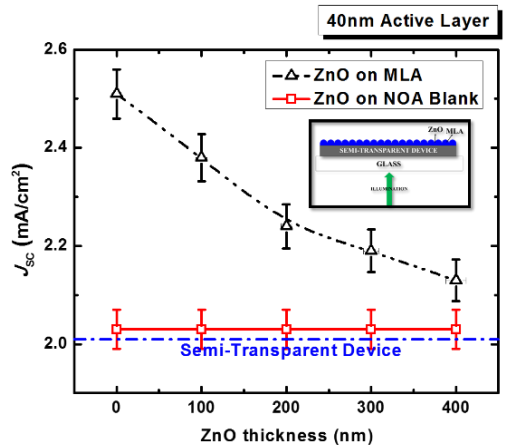

Figure 3-18. Variation in short-circuit current density, $J_{SC}$ for different ZnO thickness on MLA and NOA blank incorporated over OMO trilayer in a semi-transparent device. Figure inset shows a schematic for a device with illumination of light through glass and MLA with ZnO on OMO trilayer. The $J_{SC}$ of a baseline semi-transparent device is given as a dotted line at the bottom.

Figure 3-19 shows the results for incorporation of textured reflector (MLA/*ZnO*/Metal) with varying ZnO in 40 nm devices. It compares the enhancement in short-circuit current density, $J_{SC}$ (%) of a textured reflector (TR) with respect to a Planar

Reflector (PR) for varying ZnO thickness. It is observed that first the enhancement increases with increasing ZnO and then starts decreasing with increasing ZnO. A peak is observed around 240-260 nm. The initial increase (0-250 nm) is due to the minimization of shadowing effect and improvement in metal film coverage, which reduces the optical losses. The latter decrease (250-400 nm) is due to loss of texture (concavity) of the reflecting surface. Thus, the optimum ZnO thickness of 240-260 nm provides a balance between the two competing factors. In this work, 300 nm is chosen because of the difficulty in precise control of ZnO thickness.

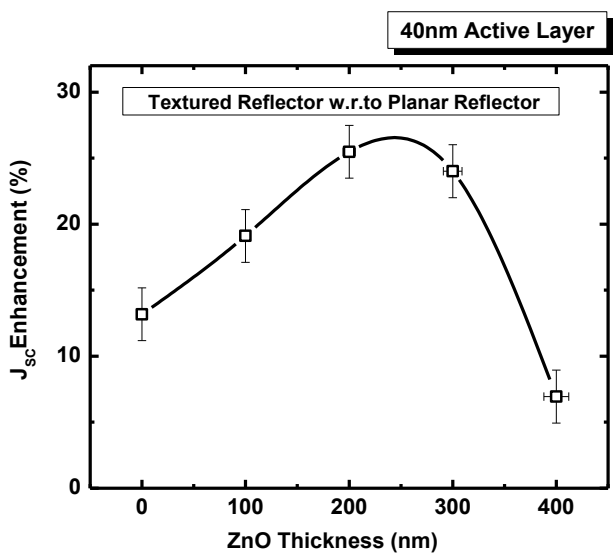

Figure 3-19. Optimization of ZnO thickness: Variation in enhancement of short-circuit current density ($J_{SC}$) for a textured reflector (w.r.to planar reflector) as a function of ZnO thickness. The textured reflector is incorporated on OMO trilayer of a semi-transparent device with 40 nm P3HT:PCBM active layer.

### 3.4.2 Enhancement Characterization

The performance of devices with textured reflector is compared to that of a planar reflector (schematic in Figure 3-17) for demonstrating the enhancement achieved and understanding the use of textured reflector.

### 3.4.2.1 Current-Voltage Characteristics

Figure 3-20(A) shows the *J-V* curves under illumination for the semi-transparent (ST) device without reflector, with planar reflector and with textured reflector for 40nm P3HT:PCBM active layer. There is not much change (<2%) in the fill factor (FF) and open-circuit voltage ($V_{OC}$) as expected. However, there is a considerable increase in the short-circuit current density ($J_{SC}$). The addition of a planar reflector increases the $J_{SC}$ by ~30%. This is because, the light which was escaping through the semi-transparent electrode before are now reflected back into the device. Addition of a textured reflector to a ST device increases the $J_{SC}$ value by ~60%, which is ~24% in comparison to a planar reflector. The additional increase is due to the increased path length and light trapping within the active layer. The power conversion efficiency also shows a similar enhancement around ~25% with respect to that of a planar reflector.

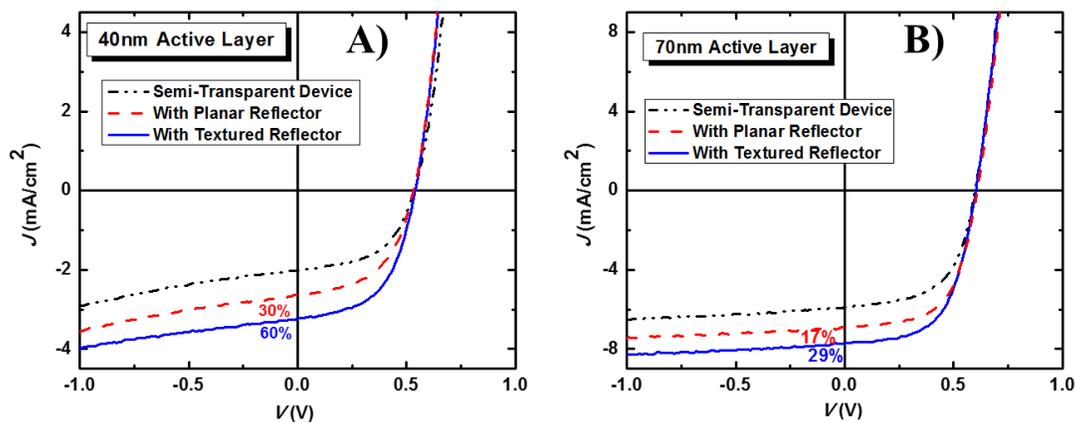

Figure 3-20. Current-Voltage (J-V) characteristics for a semi-transparent device without reflector, with planar reflector and textured reflector. P3HT:PCBM is the active layer used. A) 40 nm active layer. B) 70 nm active layer.

The study is repeated again with a 70nm active layer to understand the effect of active layer thickness. As shown in Figure 3-21, with increasing active layer thickness the amount of light transmitted decreases. This is understandable because absorption in the active layer increases with thickness as per Beer-Lambert's law. A 70nm active layer has ~13% less transmittance in comparison to a 40nm active layer.

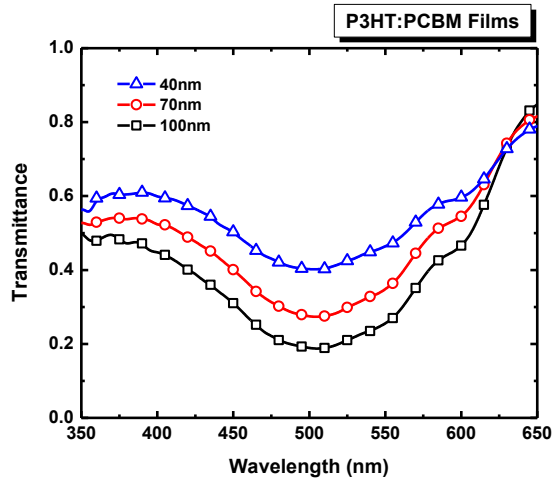

Figure 3-21. Transmittance for P3HT:PCBM films (40 nm , 70 nm, 100 nm thickness) for a wavelength range from 350 nm to 650 nm. This is the wavelength range for light absorption in P3HT:PCBM films.

Table 3-1. Device performance characteristics for incorporation of a planar reflector and textured reflector in a semi-transparent device with 40 nm and 70 nm active layer (P3HT:PCBM).

|  | 40nm Active Layer | | 70nm Active Layer | |
| --- | --- | --- | --- | --- |
|  | $J_{SC}$ (mA/cm$^2$) | $\eta_P$ (%) | $J_{SC}$ (mA/cm$^2$) | $\eta_P$ (%) |
| Semi Transparent Device | 2.0±0.1 | 0.58±0.02 | 5.9±0.1 | 2.1±0.1 |
| Planar Reflector (PR) | 2.6±0.1 | 0.75±0.03 | 6.9±0.1 | 2.5±0.1 |
| Textured Reflector (TR) | 3.2±0.1 | 0.94±0.04 | 7.6±0.2 | 2.7±0.1 |
| Enhancement (%) (TR w.r.to PR) | 24±2 | 25±3 | 10±2 | 8±2 |

Subsequently, the textured reflector would show less enhancement than 40nm active layer. Figure 3-20(B) shows the *J-V* curves for a 70nm active layer under illumination. The planar reflector shows ~17% enhancement. Whereas, the textured reflector shows ~29% enhancement with respect to a ST device. Only ~10% enhancement is achieved for textured reflector in comparison to planar reflector (as opposed to ~30% for 40nm active layer). The explanation for enhancements are similar to that of a 40nm active layer. The device performances for 40nm and 70nm active layer are consolidated in Table 3-1 for easy comparison. Thus, device absorption also plays a crucial role in the effectiveness of the textured reflector (more in Section 3.4.3).

The relative enhancements for textured reflector with respect to planar reflector as a function of active layer thickness is plotted in Figure 3-22. It is observed that enhancement decreases with increasing active layer thickness. Accordingly if the trend is extended, for a 100nm active layer which has ~9% lower transmittance than a 70nm active layer, a meagre (1-2%) or no enhancement will be observed by using textured reflector. This helps us understand that textured reflector is not effective for thicker active layers with high absorption. It has to be used in OPV devices which have high $V_{OC}$ and low absorption (low $J_{SC}$).

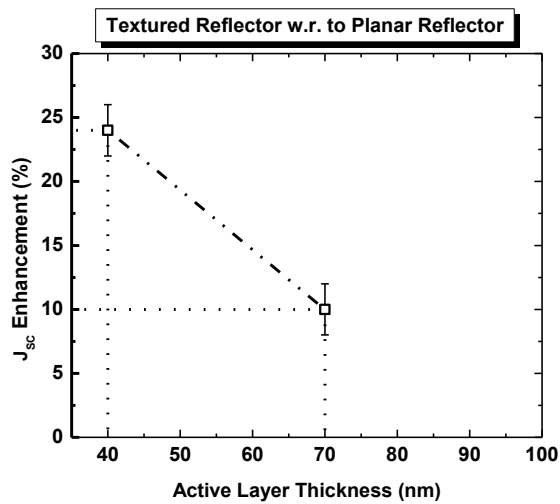

Figure 3-22. Variation in short-circuit current density ($J_{SC}$) enhancement with active layer thickness for incorporation of textured reflector. The enhancement is calculated with respect to a planar reflector.

### 3.4.2.2 External Quantum Efficiency Characteristics

The enhancement characteristics for different wavelength regimes is studied using external quantum efficiency measurement. Figure 3-23 shows the EQE of devices (70nm P3HT:PCBM active layer) with planar and textured reflector.

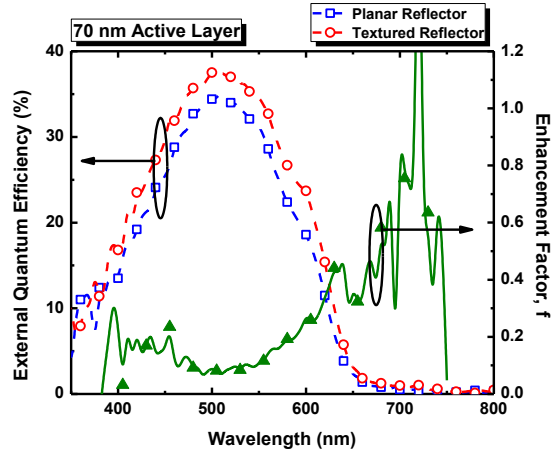

Figure 3-23. External Quantum Efficiency for devices with planar and textured reflector. The enhancement factor as a function of wavelength for textured reflector with respect to planar reflector is given by solid triangles.

As expected, device with textured reflector has higher EQE in the entire wavelength range when compared to device with planar reflector. The enhancement in EQE is only due to increase in optical absorption, since the efficiencies of exciton diffusion, charge transfer and charge collection are unaffected because of the used optical approach. The enhancement factor, $f$ can be used to understand the relative enhancements at every wavelength and is given by: $f = \frac{(EQE_{TR} - EQE_{PR})}{EQE_{PR}}$. It is shown by solid triangles in Figure 3-23 and can be observed that the enhancement factor is not uniform at all wavelengths. We can see that a lower enhancement ($f = 0.1 - 0.2$) is achieved at wavelengths (400 nm - 650 nm) where absorption is high. Whereas, a higher enhancement ($f > 0.3$) is achieved at wavelengths (650-725 nm) where the initial absorption is low. This is understandable because, for low absorption wavelengths the amount of light reflected by the textured reflector increases which results in higher enhancements. Also, the enhancement in EQE matches well with that of the J-V measurement.

### 3.4.2.3 Angle dependent Characteristics

It is essential to study the performance of solar cell with different incident angle of illumination, because solar illumination angle varies throughout the day without any additional solar tracking systems. Devices (70nm P3HT:PCBM active layer) with planar and textured reflector were used to study the angle dependency. Angle dependent measurements were made using the EQE setup described in Section 1.8.2. The wavelength of light was set at 550nm, which is close to the peak absorption wavelength for P3HT:PCBM. Then, lock-in amplifier was used to monitor the photocurrent of the devices with different tilt angle ($\theta$) varying from $0^0$ to $90^0$. Figure 3-24 shows the variation in photocurrent (@550nm) with tilt angle for devices with planar and textured reflector. It can be observed that in both these cases (PR and TR), there is a steady drop in the photocurrent with increasing tilt angle. This occurs because, the drop in absorption due to increase in surface reflectivity with tilt angle is greater than the rise in absorption due to increase in path length with tilt angle. The enhancement for TR with respect to PR is almost constant (10±2%) until tilt angle $60^0$. However beyond this, the enhancement increases and reaches up to 40% for a tilt angle of $85^0$. The higher enhancements for larger tilt angles (> $60^0$) can be attributed to a greater compensation of surface reflection by increased path length, which is possible because of the texture of reflector at the rear end. The enhancements at different tilt angles are shown by open triangles in Figure 3-24.

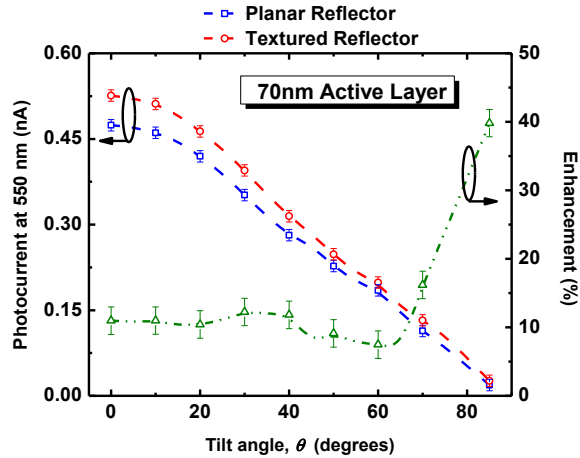

Figure 3-24. Variation of photocurrent at 550nm for different angles of light illumination. Enhancement for TR with respect to PR is shown by open triangles.

### 3.4.3 Suitability of Textured Reflector for Various Device Architectures

The textured reflector proposed here is incorporated at the rear end of the device in a superstrate configuration. The purpose of glass substrate is twofold: to hold the device and act as window for solar illumination. However a number of things should be studied carefully if this concept has to be extended to a substrate configuration, where the device would be built on a pre-fabricated textured reflector (more information in Section 5.2).

In superstrate configuration, the light reaches the textured reflector at the rear end only after the first pass through the active layer. The amount of light absorbed by the active layer and transparent electrodes are very crucial as it determines the amount of light available for reflection. If the active layer material is itself strongly absorbing around 90% of light, incorporation of TR is not meaningful. Based on the study of TR for different P3HT:PCBM thickness in the previous section, some general conclusion could be drawn based on values at wavelength with maximum absorption (~550 nm for P3HT:PCBM active layer). The textured reflector would be most beneficial for the material systems which absorb less than 50% of the light. Moderate enhancements (around 10-15%) could be achieved for materials absorbing 50-70% light. For high absorbing materials (>70%), TR is not the best of options to achieve enhancement in the performance. However as the absorption is not uniform at all wavelengths for OPVs, TR can still be used to improve the absorption at wavelength regions with low absorption.

Textured Reflector is a universal optical approach, as it can be applied to most of the material systems and doesn't affect the electrical properties of the devices. In general, it can be applied to any transparent devices with planar configurations. Thus, the potential of TR can be completely tapped by a wise choice of material system.

### 3.5 Summary

In this chapter, textured rear reflector was fabricated and demonstrated to achieve enhancements in short-circuit current and power conversion efficiencies up to 10-25% for a polymer based organic solar cell. Textured reflector were fabricated using microlens array, ZnO nanoparticles and a metal film. ZnO nanoparticles were included to minimize the shadowing effect. An optimum thickness of ZnO (~240 – 260 nm) was determined to achieve a balance between optical losses and concavity of the reflector.

The enhancement achieved is due to the increase in path length of light and light trapping within the device facilitated by the textured rear reflector. The enhancement observed is strongly dependent on the active layer absorption during the first pass and as the active layer absorption increases the effectiveness of the textured reflector decreases. The transparency of the electrode also plays a crucial role in textured reflector incorporation. The scope for further improvement of this concept is discussed in Chapter 5.

# CHAPTER 4
# IMPROVED LIGHT INCOUPLING FOR MICROLENS ARRAY INCORPORATION IN ORGANIC SOLAR CELLS

## 4.1 Introduction

Microlens array (MLA) is a promising optical management technique, which involves integration of hemispherical microlens structure. It is a very commonly used approach in organic light emitting diodes (OLEDs) to enhance the light extraction / out-coupling[57–61]. On the other hand, the use of MLA in organic photovoltaics (OPVs) have been less common and there are very few reports[44,45,62]. In OPVs, MLA is attached to the light incoupling surface and it results in altering the interaction of light with the device. It causes an overall increase in the path length within the active layer leading to an increase in short-circuit current and power conversion efficiency.

In this chapter, incorporation of microlens array on the transparent electrode is studied. Section 4.2.1 summarizes the mechanism by which MLA provides enhancement for OPV devices. In Section 4.2.2, the concept of using MLA directly over the transparent electrode is introduced and discussed. It provides an understanding about the importance of this concept and its potential to avoid optical losses. Section 4.3 includes experimental details about device fabrication (Section 4.3.1) and incorporation of microlens array (Section 4.3.2). The device performance for MLA incorporation on glass and transparent electrode are compared in Section 4.4.1. Initial results for incorporation of microlens array in ITO free devices are provided in Section 4.4.2. Finally, a summary of the discussions in this chapter is given in Section 4.5.

## 4.2 Microlens Array (MLA)

### 4.2.1 Mechanism for Enhancement using MLA in OPVs

The use of microlens array as an universal optical approach to enhance the efficiency of organic solar cells was proposed by Myers et al.[45] Understanding the mechanism by which microlens array provides enhancement for OPVs is essential in the context of this work. Hence, a brief summary about microlens in OPV is provided here.

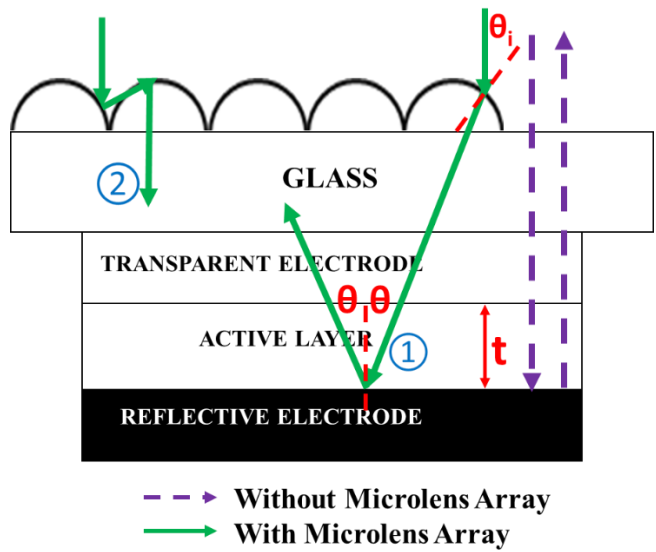

Figure 4-1. Schematic of light path in an OPV device with (continuous lines) and without (dashed lines) microlens array. The increased path length and reduction in reflection losses are denoted by 1 and 2 respectively. The active layer thickness is denoted by t and the angle between the light ray and surface normal is denoted by θ.

Microlens array is a periodic array of hemispherical lens with curved surfaces. Figure 4-1 shows a schematic of the light path in an organic solar cell with and without MLA in the light incoupling surface. Typically, the light passes twice through the active layer with a reflective electrode (dotted line in Figure 4-1). For normal incidence, the path length of light within the active layer is limited to $2t$, where $t$ is the thickness of the active layer. The unabsorbed light escapes the device after the second pass. Also, the difference in the refractive index between air ($n_a$ = 1) and glass ($n_{glass}$ ~ 1.5) causes a portion of the incident light to be reflected back. Based on Fresenl's equation, this is estimated as ~4%. Subsequently these factors result in optical losses and limit the amount of light absorbed by the active layer. The situation is considerably changed in the presence of a microlens array. As shown in Figure 4-1, the curved surface of the lens alters the path of incident light. MLA increases the optical absorption in OPVs by two ways.

Firstly, the normal incident light with the plane of device is incident at an angle due to the curvature of lens. The angle varies with the point of incidence on the surface. It increases as we move from the top of the lens to the bottom. This increased angle of incidence causes an increase in the path length (shown by #1 in Figure 4-1). With MLA, an additional angular component increases the path length of light during each pass and is given by $1/\cos\theta$, where $\theta$ is the angle made by the light with the plane of device. Thus, the total path length for two passes increases from $2t$ to $2t/\cos\theta$ in the presence of MLA.

Secondly, the light which are incident at the lens surface with very large angles are reflected into the neighboring lens (shown by #2 in Figure 4-1). Then, the light is incident at the lens with relatively lower angles and are favorable for transmission. In this way, the amount of reflection is reduced initially with the use of an array of microlenses. Thus, the microlens array accounts for enhancement in absorption within the active layer of OPV devices with increased mean path length and reduced reflection losses.

### 4.2.2 Concept of MLA on Transparent Electrode

For incorporation of MLA in OPVs, the enhancement in short-circuit current is primarily realized because of an increased path length within the active layer. In the previous works[45,62], microlens array is incorporated on the glass (also mentioned in Section 2.3.3) and a superstrate configuration is used. The refractive index of MLA used is ~1.5. However, this is not the most effective choice. As discussed above, the increase in path length is given by $1/\cos\theta$, where $\theta$ is the angle made by the light with the plane of device. A longer path length can be achieved by increasing the value of $\theta$ (i.e.) $Path\ length \propto \frac{1}{\cos\theta} \propto \theta;\ for\ 0 < \theta < 90^0$. One way to achieve a higher $\theta$ values is by increasing the refractive index of MLA ($n_{MLA}$). To understand this, a rough estimation of the angle $\theta$ for different $n_{MLA}$ can be calculated using Snell's law. For example at angle of incidence ($\theta_i$) = $45^0$, $\theta$ = $11.4^0$ for $n_{MLA}$ = 1.45 and $\theta$ = $24.3^0$ for $n_{MLA}$ = 2.00. Thus a higher refractive index MLA can be used to achieve longer path length and in turn better enhancement in $J_{SC}$.

The incorporation of high refractive index MLA ($n$ = 2) on glass is not effective because of reflection losses. The difference in the refractive index between glass ($n \approx$ 1.45) and MLA causes a portion of light (which is at angles greater than the critical angle) to be reflected back into the air. This is schematically shown in Figure 4-2(A). The critical angle for reflection depends on the refractive indices of MLA and glass. It can be calculated using Snell's law as: $sin\theta_C = n_{glass}/n_{MLA}$. The critical angle decreases as the difference in refractive indices increases. In addition to this, as light travels significant distance in the glass which is ~1-2 mm thick, some amount of light is lost because of absorption and refraction by glass.

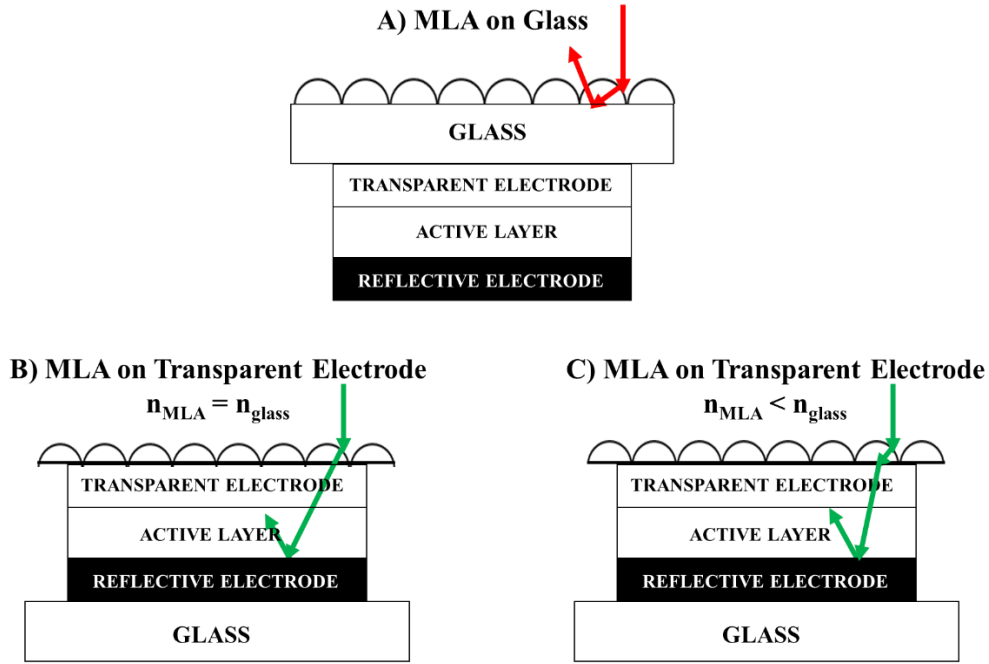

Figure 4-2. Schematic of light interactions for MLA incorporation on glass and transparent electrode. A) MLA incorporated on glass showing reflection losses at MLA / glass interface. B) MLA incorporated on transparent electrode ($n_{MLA}=n_{glass}$) without any reflection losses. C) MLA incorporated on transparent electrode ($n_{MLA}<n_{glass}$) with slightly reduced path length of light.

Accordingly the glass interface has to be avoided to improve the effectiveness of MLA. A solution to this is the incorporation of microlens array directly on the transparent electrode. To achieve this a substrate configuration has to be used with bottom reflective electrode and top transparent electrode as shown in Figure 4-2(B). Further details about device fabrication is given Section 4.3.1. When the MLA is incorporated directly on the transparent electrode (Figure 4-2(B), (C)), the reflection component because of glass interface is completely avoided. As a result, the light incoupling improves with longer path lengths and the light is completely transmitted into the device without reflection losses. The interference within the device is also improved without a glass interface.

The refractive index of the MLA can be tuned in accordance with that of the transparent electrode to achieve the best of enhancement using this approach. Ideally, the refractive indices should be matched to avoid any kind of losses and attain the maximum possible path length for light (Figure 4-2(B)). If the refractive index of MLA is lower than the refractive index of transparent electrode, reflection losses would still be avoided. However, the angular component associated with the light ray is slightly decreased due to the refraction at the MLA / transparent electrode interface (Figure 4-2(C)). This would lead to a reduction in path length of light in comparison to Figure 4-2(B).

Thus, the above explanation illustrates that MLA incorporation on transparent electrode can be effective than incorporation on glass. The explanation provided here is just qualitative. In order to attain a complete understanding about the effect of refractive index on the path length and efficiency enhancement achieved, detailed optical simulations have to be done.

### 4.3 Experimental Details

**4.3.1 Device Fabrication**

To compare and study the effect of microlens array on glass and transparent electrode, transparent devices were made. For making these devices, the available choice of transparent electrodes were Indium Tin Oxide and OMO trilayer. As OMO trilayer absorbs some light (Figure 3-4), the devices were semi-transparent. In spite of this limited transparency, these devices were used for the purpose of comparison. This is sufficient for the initial studies. However, better choice of electrodes has to be used for the final device design. Figure 4-3(A) shows the device structure of semi-transparent device with reflector deposited on the glass. This structure was used to study MLA incorporation on the OMO trilayer. Figure 4-3(B) shows the device structure of semi-transparent device with reflector on the OMO trilayer. This structure was used to study MLA incorporation on glass. Cytop interlayer was used before reflector deposition on the OMO trilayer, in order to protect the device from damage and avoid electrical shunts. The reflector used here is 100 nm aluminum. P3HT:PCBM bulk heterojunction (BHJ) of 100nm thickness was used as the active layer. Further details about fabrication of these devices are explained in Section 3.3.1.

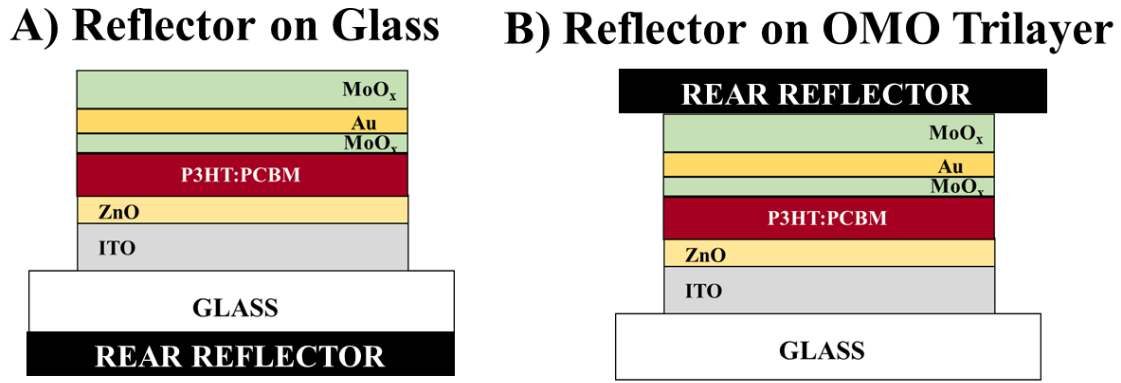

Figure 4-3. Device structure for semi-transparent device with rear reflector. A) Reflector on Glass. B) Reflector on OMO trilayer.

To implement and extend the concept of MLA incorporation on transparent electrode, ITO-free devices were made in substrate configuration as shown in Figure 4-4. ITO was replaced by 60 nm aluminum. A 40 nm ZnO was used as an electron selective layer. Then, 100 nm P3HT:PCBM BHJ was deposited as active layer. Finally, the device was completed with deposition of OMO trilayer (semi-transparent to electrode). Spin coating has to be done with care to avoid any kind of pin holes in the film due to the relatively poor wettability of ZnO on aluminum with comparison to ITO.

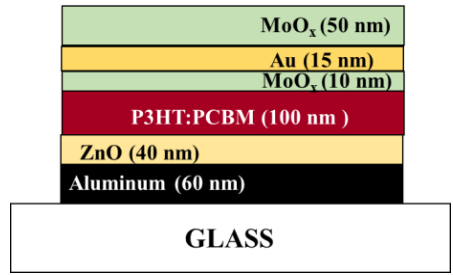

Figure 4-4. Device structure of ITO free devices (substrate configuration) with 100 nm P3HT:PCBM BHJ as the active layer.

### 4.3.2 Incorporation of Microlens Array

A detailed description about the fabrication procedure of the microlens array is provided in Section 3.3.2. Based on the above discussion, the MLA should have a refractive index ~2 for achieving the best performance. However, the microlens array used here are formed using Norland Optical Adhesive (NOA) 61, which has a refractive index ~1.58 (varies with wavelength). In spite of lower refractive index, it was used because of material limitations. This is good enough to have an initial understanding and study the potential of this approach. Further work has to be done for using high refractive index MLA and exploit its full potential (Section 5.2). When the MLA is incorporated on the OMO trilayer, care should be taken to avoid any damage to the device. Approximately 12 minutes of curing (for MLA on trilayer) and 7 minutes of curing (for MLA on glass) in 365 nm UV light has to be done before peeling of the PDMS mold.

### 4.4 Results and Discussions

### 4.4.1 Comparison of MLA on Glass and Transparent Electrode

As mentioned in Section 4.3.1, semi-transparent devices with rear reflector were fabricated to compare the effect of MLA on glass and transparent electrode. Here, the transparent electrode used for MLA incorporation is OMO trilayer. At the outset, it is important to understand that the device performance of the semi-transparent devices used here will considerably differ when illumination is through glass side and OMO trilayer side. This is solely attributed to the transparency of the OMO trilayer electrode used. Figure 4-5 shows the external quantum efficiencies (EQE) of the semi-transparent devices measured through glass side and OMO trilayer side. Also, the transmittance of ITO and OMO trilayer as a function of wavelength is shown in Figure 4-5. Compared to ITO, OMO trilayer has relatively low transmittance in the wavelength range of 350-650 nm. This low transmittance greatly reduces the EQE of the device measured through trilayer side. However, the device measured through the glass side is unaffected by this and thus a significantly higher EQE is observed in this wavelength range. For instance, at 500 nm the transmittance of ITO and trilayer is ~90% and ~65% respectively. The EQE at this wavelength range reduces from ~50% (for glass side measurement) to ~30% (for trilayer side measurement). This accounts for a lower shorter-circuit current, when the device measurements are done through the trilayer side. Increasing the transparency of the electrode used is a solution to achieve better transparent devices.

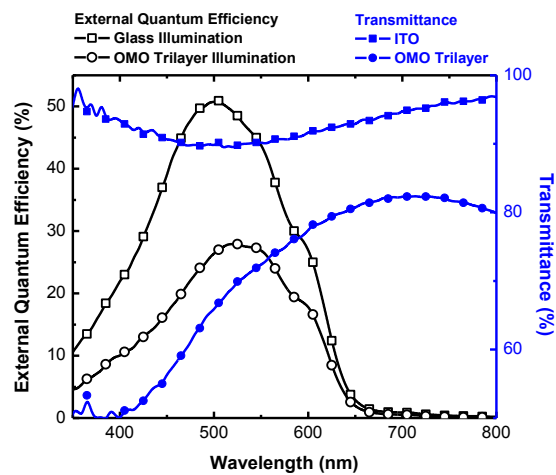

Figure 4-5. Comparison of external quantum efficiency of semi-transparent device measured through glass side and OMO trilayer side. The transparency of ITO and OMO trilayer is also shown in the figure.

In a semi-transparent device with rear reflector on the OMO trilayer, MLA was incorporated on the glass side. Figure 4-6(A) shows the J-V characteristics of this device with and without microlens array. It is observed that only ~6% enhancement is observed in the short-circuit current with the use of MLA on the glass ($J_{SC}$ is 6.9±0.2 mA/cm$^2$ without MLA and 7.4±0.2 mA/cm$^2$ with MLA). The enhancement in power conversion efficiency is approximately same (~5%) because there is very minimal change in the open-circuit voltage and fill factor.

On the other hand, in a semi-transparent device with rear reflector on the glass, MLA was incorporated on the OMO trilayer. Figure 4-6(B) shows the J-V characteristics of this device with and without microlens array. It is observed that short-circuit current increases by ~28% with the incorporation of MLA on the OMO trilayer ($J_{SC}$ is 4.3±0.1 mA/cm$^2$ without MLA and 5.5±0.1 mA/cm$^2$ with MLA). The increase in power conversion efficiency is also ~30%. Here it can be seen that without MLA, the $J_{SC}$ for measurement through trilayer side (4.3±0.1 mA/cm$^2$) is significantly lower than the $J_{SC}$ for measurement through glass side (6.9±0.2 mA/cm$^2$). This is in accordance to the discussion above related to the transmittance of OMO trilayer. For convenience, the data is consolidated in Table 4-1.

Based on the results above, a significant difference in the enhancement can be observed for MLA incorporated on glass side (~6%) and trilayer side (~28%). As all other factors in the comparison are same, the differences in enhancement is a clear indication of additional optical losses for MLA incorporated on glass with respect to that of the MLA incorporated on trilayer. The source for these optical losses may be as discussed in Section 4.2.2.

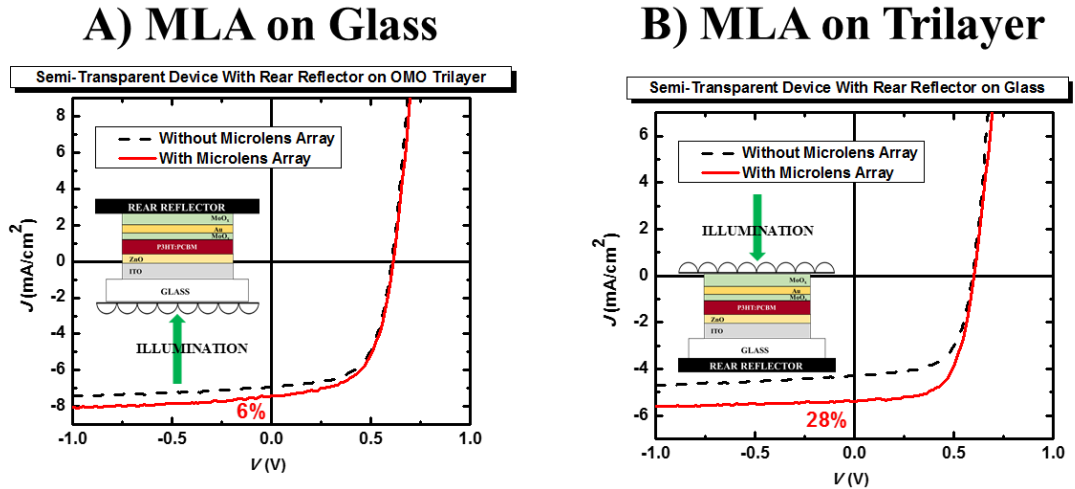

Figure 4-6. Current-Voltage characteristics for semi-transparent devices with rear reflector. A comparison is provided for with and without microlens array (MLA). A) MLA incorporated on the glass. B) MLA incorporated on the OMO trilayer. (Note: Figure insets shows the device structure along with the direction of illumination)

Thus, it can be concluded that incorporation of MLA on OMO trilayer (semi-transparent electrode) is much effective than incorporation of MLA on glass.

Table 4-1. Device performance characteristics of a semi-transparent device with rear reflector along with MLA incorporation on glass and OMO trilayer.

|  | Semi-Transparent device with Rear Reflector (100nm Active Layer) | | | |
|---|---|---|---|---|
|  | Microlens on Glass | | Microlens on OMO Trilayer Electrode | |
|  | $J_{SC}$ (mA/cm$^2$) | $\eta_P$ (%) | $J_{SC}$ (mA/cm$^2$) | $\eta_P$ (%) |
| Without Microlens Array | 6.9±0.2 | 2.6±0.1 | 4.3±0.1 | 1.6±0.1 |
| With 4μm Microlens Array | 7.4±0.2 | 2.7±0.1 | 5.5±0.1 | 2.1±0.1 |
| Enhancement (%) | 6±2 | 5±2 | 28±1 | 30±1 |

### 4.4.2 ITO-Free Devices with Microlens Array

Indium Tin Oxide (ITO) is the most commonly used transparent electrode in organic photovoltaics. However, this is not viable from an economic viewpoint. ITO uses expensive rare metal indium as a raw material and requires high energy intensive manufacturing process. As a result, ITO is very expensive and it greatly affects the commercialization of OPV technology. Lots of efforts are being made in the development of ITO-free devices and achieve high efficiencies[25,63–65].

The efficiencies of ITO-free devices can be enhanced through the optical approach by using the microlens array (MLA). As MLA is a universal optical approach, it can be used in any kind of device architecture and material systems. The concept of MLA incorporation on transparent electrode (discussed in Section 4.2.2) is very relevant and applicable to ITO-free devices. Hence, an attempt is made here to study the incorporation of MLA in ITO-free devices.

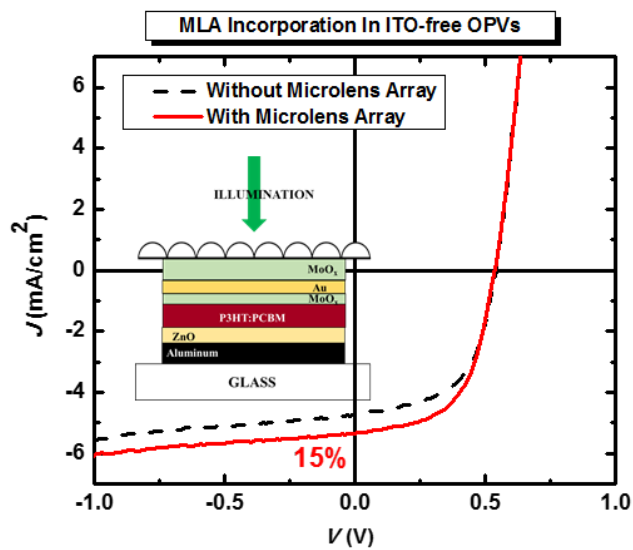

Figure 4-7. Current-Voltage characteristics of ITO-free device with and without microlens array. The figure inset shows the device structure and illumination direction.

The ITO-free devices used here are in substrate configuration, where aluminum on glass is used as the bottom electrode and OMO trilayer electrode on the top is used for illumination. Further details about the device fabrication is given in Section 4.3.1. Since the light illumination is through OMO trilayer, the above discussed concept can be extended here. Figure 4-7 shows the J-V characteristics of an ITO-free device, with and without MLA. It is seen that with a microlens array, short-circuit current ($J_{SC}$) increases from 4.7±0.2 mA/cm² to 5.4±0.2 mA/cm², which is approximately a ~15% enhancement. A similar enhancement (~14%) is observed in the power conversion efficiency because there are only negligible changes in open-circuit voltage and fill factor. The enhancement can be attributed to the MLA though mechanisms discussed in Section 4.2. The data is consolidated in Table 4-2.

Table 4-2. Device performance characteristics of ITO-free devices with and without microlens array.

|  | $J_{SC}$ (mA/cm²) | $\eta_P$ (%) |
|---|---|---|
| Without Microlens Array | 4.7±0.2 | 1.4±0.1 |
| With 4µm Microlens Array | 5.4±0.2 | 1.6±0.1 |
| Enhancement (%) | 15±2 | 14±1 |

The device structure is Figure 4-6(B) and Figure 4-7 are very similar. The only difference is the position of aluminum within the device. In Figure 4-6(B), there is an additional glass interface between the aluminum and device. Whereas, there is no kind of glass interface between aluminum and the device in Figure 4-7. This causes a significant difference in the interference effect within the device leading to the changes in the short-circuit current. A higher short-circuit current of 4.7±0.2 mA/cm² is observed when there is no glass interface (ITO-free device). However, short-circuit current decreases to 4.2±0.1 mA/cm², when the glass interface is present (semi-transparent device with rear reflector on glass). This signifies that due to the glass interface, a poor interference occurs causing less light to be absorbed within the active layer. However with a microlens array, the short-circuit current achieved is very similar in both cases (5.4±0.2 mA/cm² and 5.5±0.1 mA/cm²). Therefore, a higher enhancement (~28%) is observed by using MLA in the case with glass interface due to the presence of more unabsorbed light. Whereas, the enhancement is lowered (~15%) because of the better baseline for ITO-free devices (as a result of strong interference effect). This accounts for the difference in enhancement values.

To further understand the enhancement observed in ITO-free device, external quantum efficiency (EQE) measurements were made. Figure 4-8 shows the EQE of ITO-free devices with and without microlens array. As expected, with MLA incorporation there is an increase in the EQE over entire wavelength range. The enhancement factor, *f* can be used to understand the relative enhancements at every wavelength. It is calculated using the expression, $f = \frac{(EQE_{MLA} - EQE_{No\ MLA})}{EQE_{No\ MLA}}$. It is shown by the blue line in Figure 4-8 and can be observed that the enhancement factor is not uniform at all wavelengths. We can see that a lower enhancement (*f* = 0.1 – 0.2) is achieved at wavelengths (400 nm - 650 nm) where absorption is high. Whereas, a higher enhancement (f = 0.3 - 1) is achieved at wavelengths (650-725 nm) where the initial absorption is low. This is true for any optical approach because less absorbed light gets chance to pass through active layer several additional times in comparison to the highly absorbed light, leading to a higher enhancement in regions with low absorption. Also, this enhancement matches well with that of the J-V measurement.

Consequently, it can be confirmed that the microlens array causes an increase in the amount of light absorbed by ITO-free devices resulting in better short-circuit current and power conversion efficiency. Still, the enhancement achieved can be improved by matching the refractive index of MLA with that of the transparent electrode because of the reasons discussed in Section 4.2.2. Thus, the use of MLA is a promising approach for improvement of ITO-free devices.

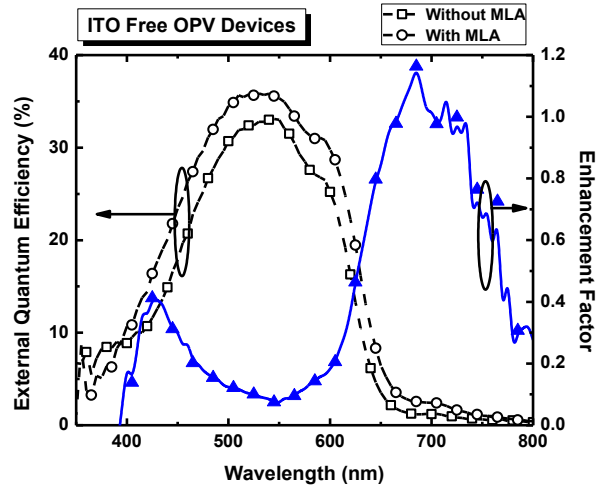

Figure 4-8. External Quantum Efficiency for ITO-free devices with and without microlens array (MLA). The blue line denotes the enhancement factor as a function of wavelength for the MLA incorporation.

## 4.5 Summary

The increase in path length achieved using MLA can be improved by increasing the refractive index of MLA. The presence of additional glass interface with high refractive index MLA results in optical losses through reflection. In this chapter, incorporation of microlens array on the transparent electrode is suggested as solution to avoid optical losses and improve the effectiveness of MLA incorporation in OPVs. Using a semi-transparent device with a rear-reflector, it is shown that MLA incorporation on glass provides ~6% enhancement and MLA incorporation on OMO trilayer provide ~28% enhancement. Also, the MLA incorporation is demonstrated for ITO-free devices. It is observed that glass interface adversely affects the interference and in turn reduces the initial absorption within the device. The gain from interference increases the baseline device performance for ITO-free device. As a result, slightly lower ~15% enhancement is observed by using MLA on OMO trilayer in ITO-free devices.

The work presented in this chapter are just initial results to understand the advantages and potential of MLA incorporation on transparent electrode without any glass interface in between. However, detailed optical simulations have to be done to explore this concept completely. The steps / direction for further work is discussed in Chapter 5.

# CHAPTER 5
# CONCLUSIONS AND FUTURE WORK

## 5.1 Conclusion

Currently, most of the research efforts for improving the efficiency of organic solar cells are towards development of new materials and device structures. Relatively few works are focusing on the optical management techniques, which is a promising approach. Hence, this thesis was centrally focused on developing optical structures for OPVs. The work presented in this thesis can be divided into two parts. The first part is related to the development and incorporation of a textured rear reflector. The second part describes an idea for improving the light incoupling of microlens array incorporated in organic solar cells.

### 5.1.1 Textured Rear Reflector

In Chapter 3, the rear reflector was textured to achieve an increased path length within the active layer and facilitate light trapping in OPV devices. Microlens array (MLA) formed by soft lithography was used to fabricate the textured reflector (TR). Metal film (Ag) was deposited on MLA to form the TR, reciprocating the texture of MLA. Zinc oxide (ZnO) nanoparticles were used to avoid the shadowing effect which occurred during deposition of metal film. Spin coating was used for deposition of ZnO over MLA. Approximately 250-300 nm was determined as the optimum ZnO thickness to achieve a balance between the optical losses and concavity of the reflector.

Polymer (P3HT:PCBM) based semitransparent solar cells were used to demonstrate the efficiency enhancement of TR with respect to a planar reflector (PR). A superstrate configuration was used, where the light passed through the active layer before reaching the reflector at rear end. Around 10-25% enhancements in $J_{SC}$ and power conversion efficiency were observed for different active layer thicknesses through incorporation of TR. The efficiency enhancement was observed to be decreasing with increase in the light absorption by active layer during the first pass. Thus, the rear TR can be used as an effective light management technique in OPV devices which have low $J_{SC}$ (low absorption) and high $V_{OC}$ values.

### 5.1.2 Incorporation of MLA on Transparent Electrode

In Chapter 4, the concept of incorporating MLA on transparent electrode was explored to improve the enhancement achieved when compared to incorporation on glass. Using this approach, the path length of light can be further enhanced by increasing the refractive index of MLA. The presence of glass interface between MLA and device would result in reflection losses. However, an improved light incoupling and longer path length is achieved when MLA is incorporated on transparent electrode. Semi-transparent device with rear reflector (100nm Aluminum) was used to study the enhancement characteristics.

When MLA was incorporated on glass, ~6% enhancement in $J_{SC}$ and PCE was observed. Whereas, a much higher enhancement of ~28% was achieved when MLA was incorporated on OMO trilayer electrode. Also, it was found that a glass interface between the reflector and device caused a reduction in short-circuit current due to poor interference effect within the active layer. This improved the baseline ITO-free device and a slightly lower (~15%) enhancement was achieved with MLA incorporation on OMO trilayer. Thus, the initial results attained here displays the merits for MLA incorporation on transparent electrode rather than glass.

## 5.2 Future Work

Optical management is a promising approach for improving the efficiencies for OPVs. Despite very few reports are available till date. Hence, in future significant amount of work has to be focused on developing new optical management techniques. In this section, some suggestions for further work related to textured reflector and MLA incorporation are provided.

### 5.2.1 Textured Reflector

In Chapter 3, textured reflector was implemented in a superstrate configuration. The design of TR can further be made effective by increasing the refractive index of MLA greater than the filling material used. Furthermore, the concept of TR can be extended to a substrate configuration. Figure 5-1 shows a possible substrate configuration with a textured reflector. In this configuration, the entire device is built on the TR and the curvature is reciprocated in the subsequent layers. The texture over the entire device would greatly facilitate the trapping of incoming light. As a result of this, much better optical absorption could be achieved. This would be a simpler and inexpensive approach to achieve texture in the device. The continuity of TR surface has to be understood initially to develop this structure. Vacuum annealing has to be used instead of thermal annealing for device fabrication (to avoid the damage of TR).

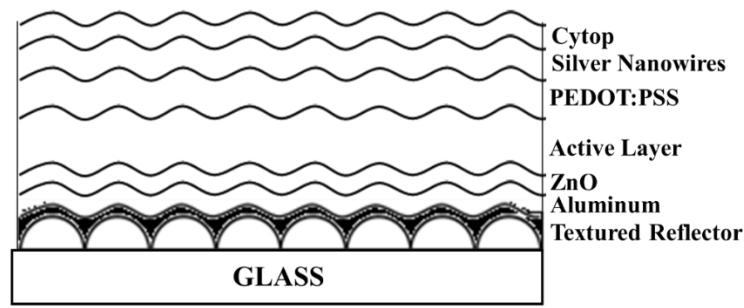

Figure 5-1. Possible substrate configuration with texture reflector.

In order to extend the concept of TR to small molecule devices, a transparent device is required. To achieve this, inverted small molecule device structures[66] similar to polymer based OPVs have to be developed, so that ITO and OMO trilayer can be used as two transparent electrodes. One possible structure is: Glass / ITO / ETL / Active Layer / HTL / OMO Trilayer. The selection of proper electron transport layer (ETL) is the challenging part. Initial studies were done with $Cs_2CO_3$[67] and $ZnO$[68] for ETL. However, the devices were leaky and no success was attained in fabricating a working device. Additionally, efforts have to be taken to develop better transparent electrodes.

**5.2.2 Microlens Array**

As discussed in Chapter 4, the best of enhancement for MLA can be achieved when the refractive index is matched with that of the transparent electrode (n ≈ 2). However, it is not practical to have a transparent polymer with such a high refractive index. High-index inorganic nanoparticles (i.e.) $TiO_2$ could be mixed with a conventional transparent polymer such as NOA, to increase the refractive index. The particles with diameters much less than the wavelength of light will only increase the effective refractive index and not scatter light. The method is limited by the processability of the high index nanoparticles.

Also, a robust simulation method has to be developed to accurately couple millimeter-scale substrate and micrometer-scale lens dimensions with nanometer-scale device active layer thickness. Finite difference time – domain (FDTD) calculation is one such method and numerous commercial software packages are available to perform this.

# LIST OF REFERENCES

# BIOGRAPHICAL SKETCH

Adharsh Rajagopal grew up in Erode, a town located in southern India. He completed his high school in 2008. He has shown a thirst for learning during his schooling and excelled in studies. Through a National Level entrance examination, he entered into one of the premier institutions in India, M. A. National Institute of Technology - Bhopal and pursued his undergraduate studies. He graduated with Bachelor of Technology degree in Material Science and Metallurgical engineering in 2012. During his undergraduate studies, he received "Silver Medal" for his academic distinction.

The appetite for higher education and research made him to search for opportunities worldwide. He went to University of Florida, Gainesville for his Master of Science degree in Material Science and Engineering. During his master's degree, he worked in the research group of Professor Jiangeng Xue and completed his thesis on 'Optical Management Techniques for Organic Solar Cells'. His academic excellence during graduate studies has earned him a membership in Tau Beta Pi, an engineering honor society. He graduated with Master of Science degree in August 2014.

He would be joining University of Washington, Seattle in fall 2014 to pursue his doctoral studies in Materials Science and Engineering and continue his research in solar cells.